\documentclass{article}

\pdfoutput=1

\usepackage[fleqn]{amsmath}
\usepackage{cite,latexsym,amssymb,bm,pbox}
\usepackage{bbm}

\newcommand{\diagram}[2][{}]{\pbox{\textwidth}{\includegraphics[#1]{{#2}}}}

\usepackage{graphicx}
\usepackage{color}
\usepackage{graphics}
\usepackage{float}
\usepackage{afterpage}
\usepackage[latin1]{inputenc}
\usepackage{caption}
\usepackage{subfig}

\usepackage[pdftex,backref,colorlinks]{hyperref}
\usepackage[figure]{hypcap}
\usepackage[letterpaper]{geometry}
\newcommand{\tr}{{\rm tr}}               
\newcommand{\cut}{\hspace{1.5mm}}
\newcommand{\tol}[1]{$\overline{\text{#1}}$}

\begin{document}

\noindent{\Large\bf New observables to test the Color Glass Condensate beyond the large-$N_c$ limit}

\vspace{0.5cm} {\large { Cyrille Marquet$^1$ and Heribert Weigert$^2$} }
\\\vspace{0.2cm}\\
{\small {
    $^1$ Institut de Physique Th\'eorique, CEA/Saclay, 91191 Gif-sur-Yvette cedex, France \\
    $^2$ Department of Physical Sciences, University of Oulu, P.O. Box 3000, FI-90014 Oulu, Finland} \\
}

\vspace{.2cm}
\noindent\begin{center}
\begin{minipage}{.92\textwidth}
  {\sf The JIMWLK framework offers a powerful tool to calculate the energy
    dependence of QCD observables at high energies. Despite a growing number
    of observables considered for phenomenological analysis, few features of
    JIMWLK evolution beyond its evolution speed are yet well constrained by
    experiment. We argue that meson production cross-sections have the
    potential to provide qualitatively new insights and allow to address
    issues both beyond the large-$N_c$ limit and at higher twist. These
    cross-sections generically contain four point functions whose evolution is
    shown to follow from the JIMWLK framework. The Gaussian truncation is used
    to provide an efficient and practical means of calculating the evolution
    of four point correlators beyond the large-$N_c$ limit.
}
\end{minipage}
\end{center}
\vspace{1cm}

\section{Introduction}
\label{sec:introduction}

Hard processes in hadronic collisions, which resolve the partonic structure of
hadrons, are well described by the leading-twist approximation of QCD. In this
weak-coupling regime, partons in the hadronic wave function scatter
independently, this is the essence of collinear factorization. However, since
the parton densities grow with decreasing energy fraction $x,$ the hadronic
wave function also features a low-$x$ part where the parton density has become
large and partons scatter coherently, invalidating the leading-twist
approximation. This weak-coupling regime, where non-linearities are important,
is called saturation, and it can be probed at high-energies since increasing
the energy of a collision allows to probe lower-energy partons.

The Color Glass Condensate (CGC) has been proposed as the effective theory to
describe this small-$x$ part of the hadronic wave function in
QCD~\cite{Gribov:1984tu, Mueller:1986wy, Mueller:1994rr, Mueller:1994jq,
  Mueller:1995gb, McLerran:1993ka, McLerran:1994ni, McLerran:1994ka,
  McLerran:1994vd, Kovchegov:1996ty, Kovchegov:1997pc, Jalilian-Marian:1997xn,
  Jalilian-Marian:1997jx, Jalilian-Marian:1997gr, Jalilian-Marian:1997dw,
  Jalilian-Marian:1998cb, Kovner:2000pt, Weigert:2000gi, Iancu:2000hn,
  Ferreiro:2001qy, Kovchegov:1999yj, Kovchegov:1999ua, Balitsky:1996ub,
  Balitsky:1997mk, Balitsky:1998ya, Iancu:2003xm, Weigert:2005us,
  Jalilian-Marian:2005jf,Gelis:2010nm}. Rather than using a standard
Fock-state decomposition, it is more efficient to use collective degrees of
freedom, more adapted to describe the behavior of the small-$x$ partons, which
are mainly gluons. The CGC approach uses classical color fields: the
long-lived large-$x$ partons are represented by a strong color source
$\rho\!\sim\!1/g_S$ which is static during the lifetime of the short-lived
small-$x$ gluons, whose dynamics are described by a color field ${\cal
  A}\!\sim\!1/g_S.$ The color source distribution $W_Y[\rho]$ depends on the
rapidity separation $Y=\ln(1/x)$ between the source and the field, and contains
information on parton densities and parton correlations. These color field
configurations are naturally probed in high energy collisions with a rapidity
separation of $Y\sim\ln(s)$ between projectile and target.

Using a ``dilute'' projectile to probe the dense target, it is possible to
expand the projectile into its leading Fock state components which then
interact eikonally with the target field, due to the high energy in the
scattering event that allows to justify a no-recoil approximation. In this
limit, the interaction with the color charge densities $\rho$ induce a Wilson
line factor $U_{\bm x}[\rho]$ along the world-line of each of the projectile
constituents which, due to their high longitudinal momentum, penetrate the
target field at a fixed transverse position $\bm x$. This approximation is
tantamount to neglecting a part of the higher-twist contributions not relevant
at small $x$. Since the charge densities enter only via these eikonal factors,
one may change variables and replace the $\rho$-distribution $W_Y[\rho]$ by a
Wilson-line distribution $Z_Y[U]$.

When using this picture to compute the \emph{total} cross-section in deep
inelastic scattering (DIS), the arbitrary separation $Y$ should be thought of
as a factorization ``scale.'' Requiring that the total cross-section is
independent of the choice of $Y$, a nonlinear, \emph{functional}
renormalization group equation for $Z_Y[U]$ (or equivalently $W_Y[\rho]$) can
be derived, the Jalilian-Marian-Iancu-McLerran-Weigert-Leonidov-Kovner-
(JIMWLK-) equation. Explicit expressions exist in the leading-logarithmic
approximation which resums powers of $\alpha_S\ln(1/x)$, supplemented by
running coupling corrections~\cite{Gardi:2006rp,HW-YK:2006}. The remaining
NLO-contributions can be found in a series of papers by Balitsky and
Chirilli~\cite{Balitsky:2008zz, Balitsky:2008rc, Balitsky:2009xg,
  Balitsky:2009yp}. However, to date they are not translated into generic
JIMWLK-language that would apply to arbitrary $n$-point functions, instead
they are given as corrections to \emph{dipole} correlators.

Equivalently, the JIMWLK equation can be cast into a coupled hierarchy of
equations for $n-$point Wilson-line correlators, the Balitsky
hierarchies~\cite{Balitsky:1996ub, Balitsky:1997mk,Balitsky:1998ya}. If the
lowest order Fock-state of the projectile is a $q\Bar q$ pair, like in DIS at
HERA, the corresponding total cross-section is driven (at zeroth order) by a
two-point function or dipole correlator $\langle\Hat S_{\bm x\bm y}^{q\Bar
  q}\rangle_Y:=\langle\tr(U_{\bm x} U^\dagger_{\bm y})/N_c\rangle_Y$, with the
average taken with the distribution $Z_Y[U]$.

Most of the phenomenology uses a mean-field approximation which significantly
simplifies the high-energy QCD evolution: it reduces the hierarchy to a
non-linear equation for the two-point function, the Balitsky-Kovchegov (BK)
equation~\cite{Balitsky:1996ub,
  Balitsky:1997mk,Balitsky:1998ya,Kovchegov:1999yj, Kovchegov:1999ua}. This
equation reduces to the BFKL equation when the amplitude is small, and
contains saturation effects as the amplitude reaches unity. Although this
equation is not exact, it has the advantage of being a closed equation for the
dipole scattering amplitude. In addition, closer inspection, both numerically
and semi-analytically~\cite{Rummukainen:2003ns,Kovchegov:2008mk}, reveals that
the difference between BK and JIMWLK solutions for $\langle\Hat S_{\bm x\bm
  y}^{q\Bar q}\rangle_Y$ is small -- much smaller than the order $1/N_c^2$
difference expected from a superficial analytical argument. For the energy
dependence of the \emph{total} cross-section, Next-to-Leading-Order (NLO)
corrections are much more important: scale-invariance breaking
running-coupling contributions~\cite{Gardi:2006rp, Kovchegov:2006vj,
  Balitsky:2006wa} change exact scaling behavior with an energy dependent
saturation scale $Q_s(Y)$ at asymptotically high energies into pseudo-scaling,
which is reached much earlier. Together with the remaining conformal
corrections at NLO~\cite{Balitsky:2008zz} evolution rates are drastically
reduced. This is crucially important for successful
phenomenology~\cite{Albacete:2007yr, Albacete:2007sm, Albacete:2009fh,
  Weigert:2009zz, HW-DIS2009, Albacete:2010bs}.

There are other observables in DIS for which it has been shown that the
$Y$-evolution is given by the same JIMWLK equation, through more complicated
correlators~\cite{Kovchegov:1999ji, Hentschinski:2005er, Hatta:2006hs,
  Kovner:2006ge}.  For instance, diffractive structure functions can be
computed in the CGC framework with the same level of success as inclusive
ones~\cite{Kovchegov:1999kx, Marquet:2007nf, Weigert:2009zz, HW-DIS2009}.
Diffractive gluon production has also been
investigated~\cite{Kovchegov:2001ni, Marquet:2004xa, GolecBiernat:2005fe} as
well as semi-inclusive DIS~\cite{Marquet:2009ca}.  Exclusive vector-meson
production was also considered to study how a finite momentum transfer affects
the way the $Y$-dependence predicted by evolution equations is mapped onto the
actual energy dependence of cross-sections~\cite{Marquet:2005qu,
  Marquet:2005zf, Marquet:2007qa}.

Despite this recent extension of scope beyond total cross-sections,
phenomenological treatments constrain the initial conditions to JIMWLK
evolution only in a relatively mild manner -- evolution ``speed'' is most
tightly constrained, precise details of correlator shapes such as UV anomalous
dimensions only meet weak experimental tests. This is probably best
illustrated by the fact that the HERA-fits of~\cite{Weigert:2009zz,
  HW-DIS2009} use correlators in the pseudo-scaling region of evolution while
the authors of~\cite{Albacete:2009fh} base their analysis on a pre-asymptotic
initial condition. The calculational tools and cross-sections suggested here
are intended as a starting point to begin exploring such issues.

It is therefore desirable to broaden the scope of observables and search for
quantities that directly probe subtler properties of correlators and at the
same time remain calculable in a practical sense, preferably in a suitable
truncation of full JIMWLK to sidestep the numerical obstacles posed by a full
JIMWLK simulation.

The selection of suitable observables is not a trivial task. To focus on yet
unconstrained features of JIMWLK evolution and its initial condition, one must
by necessity resort to more and more differential observables. This must be
done in a way that avoids infrared problems that would invalidate a
perturbative treatment via an evolution equation. At the same time one would
like to limit the amount of complexity, both to keep cross-sections large
enough and to keep numerical effort down: As will become obvious from the
example considered below, the price to be paid for more detailed information
is the appearance of more complicated $n$-point functions in the analytical
expressions for the cross-section, whose correspondingly involved
configuration space structures result in costly, since numerically delicate,
integrals.

The discussion below will mainly focus on meson production
cross-sections. They serve to illustrate how to calculate the energy
dependence of both momentum-transfer- ($t$-) dependent and $t$-integrated
cross-sections with specific restrictions on the final state. Properly chosen,
such restrictions can serve as a filter for information that is unaccessible
in the total cross-section. The treatment provides a model for other
observables.

A central part of this paper is played by a set of diffractive observables in DIS
--based on the target dissociative part of vector-meson production or deeply
virtual Compton scattering (DVCS)-- that have the potential to expose
unexplored features of JIMWLK evolution without excess complexity. As will be
explained in the text, these observables directly probe the correlator
difference~\cite{Dominguez:2008aa}
\begin{align}
  \label{eq:corr-diff-intro}
  \langle\Hat S_{\bm y'\bm x'}^{q\Bar q}\Hat S_{\bm x\bm
  y}^{q\Bar q}\rangle_Y -\langle \Hat S_{\bm y'\bm x'}^{q\Bar
  q}\rangle_Y\langle\Hat S_{\bm x\bm y}^{q\Bar q}\rangle_Y
\ ,
\end{align}
a slight generalization of the ``correlator factorization violations''
extensively discussed in~\cite{Kovchegov:2008mk}. Such correlators strictly
vanish in the BK-mean-field approximation (due to $1/N_c$ suppression of their
leading contributions), and are also twist suppressed compared to the total
cross-section (they require the exchange of at least four gluons between
projectile and target in the $t$-channel). The corresponding cross-sections
are necessarily small, but should probe features of JIMWLK evolution that
remain invisible in the global observables considered to date. Our main
achievement is the calculation of the correlator
difference~\eqref{eq:corr-diff-intro} (in the Gaussian truncation of
JIMWLK-evolution introduced next), which directly enters the formulation of
the diffractive dissociation cross-section~\eqref{eq:diff-diss}.  Indeed,
thanks to the hard scale provided by the photon virtuality, we are able to
calculate in QCD (in the high-energy limit), the diffractive dissociation
introduced by Good and Walker in the context of soft
diffraction~\cite{Good:1960ba}.

Since~\eqref{eq:corr-diff-intro} vanishes identically in the BK-approximation,
it becomes imperative to find an approximation capable of capturing the
relevant contributions.  By adopting the Gaussian truncation
(GT)~\cite{Kovner:2001vi,Weigert:2005us,Kovchegov:2008mk} of JIMWLK-evolution
in place of the BK-approximation, it turns out to be possible to efficiently
calculate the energy dependence of both two- and four-point correlators in
these observables. In the large-$N_c$ limit, the Gaussian truncation reduces
to the BK truncation, but in general it allows to calculate
correlator-factorization violations.  Instead of using $\langle \Hat S_{\bm
  x\bm y}^{q\Bar q}\rangle_Y$ directly as its degree of freedom, the GT uses a
two-gluon $t$-channel exchange correlator ${\cal G}_{Y,\bm{ x y}}$ that enters
the $q\Bar q$ correlator as $\langle \Hat S_{\bm x\bm y}^{q\Bar q}\rangle_Y =
e^{-C_{\text{f}} {\cal G}_{Y,\bm{ x y}}}$. Unlike the BK-approximation it can
naturally be used to parametrize more complicated $n$-point functions in a
consistent manner. This is used below to derive an expression for the four
Wilson-line correlator $\langle\Hat S_{\bm y'\bm x'}^{q\Bar q}\Hat S_{\bm x\bm
  y}^{q\Bar q}\rangle_Y$ (including its $Y$-dependence) in the Gaussian
truncation.

The dynamical content driving the energy dependence in the GT approximation is
in fact identical to that of the BK approximation~\cite{Kovchegov:2008mk}. As
with the BK approximation, its evolution equation emerges directly from the
JIMWLK equation and can therefore be extended to (in principle) arbitrary loop
order. It is, however, possible 
to systematically extend the treatment by including genuine multi-$t$-channel
gluon-correlations into the formalism, building on the non-abelian
exponentiation theorem~\cite{Gatheral:1983cz, Frenkel:1984pz, Laenen:2008gt}.

The derivation of evolution equations for the various cross-sections
considered below remains at a one loop level. This is done in part to keep the
arguments compact. A full treatment at NLO would first require a translation
of the results of Balitsky and Chirilli into JIMWLK-language, a task left for
a separate publication.

The plan of the paper is as follows. Section~\ref{sec:evolution-dis-cross},
outlines the proof at leading order of how JIMWLK-evolution determines the
small-$x$ dependence of certain diffractive cross-sections in DIS, including
vector-meson production. The calculation clarifies how, four-point Wilson line
correlators enter the more differential cross-section in addition to the
two-point functions already present in the total cross-section. The
generalization to NLO is discussed qualitatively.
Section~\ref{sec:corr-gauss-trunc} serves to recall the Gaussian truncation
approximation of JIMWLK. It is contrasted with another extension beyond the
BK-approximation to JIMWLK evolution as advocated in~\cite{Janik:2004ts,
  Janik:2004ve}. A re-derivation of the 2-point function $\langle\Hat S_{\bm
  x\bm y}^{q\Bar q}\rangle_Y$ in the Gaussian truncation provides the starting
point to address the case of non-trivial higher-$n$-point correlators -- an
expression for $\langle\Hat S_{\bm y'\bm x'}^{q\Bar q}\Hat S_{\bm x\bm
  y}^{q\Bar q}\rangle_Y$ is explicitly worked out in
Section~\ref{sec:quadr-evol-gauss}. Section~\ref{sec:meas-corr-fact-directly}
finally discusses how, in diffractive vector-meson production or deeply
virtual Compton scattering (DVCS), the target-dissociating part of the
cross-section is related to the correlator difference $\langle\Hat S_{\bm
  y'\bm x'}^{q\Bar q}\Hat S_{\bm x\bm y}^{q\Bar q}\rangle_Y -\langle \Hat
S_{\bm y'\bm x'}^{q\Bar q}\rangle_Y\langle\Hat S_{\bm x\bm y}^{q\Bar
  q}\rangle_Y$ of~\eqref{eq:corr-diff-intro}. Section~\ref{sec:conclusions}
attempts to put the results into perspective with a review of the tools
developed thus far and a short discussion of systematic improvements left for
future study.

\section{Evolution of DIS cross-sections with restrictions on the projectile
  final state}
\label{sec:evolution-dis-cross}

As already indicated in the introduction, all cross-sections in the high
energy limit of the CGC framework will be expressed in terms of Wilson-line
operators that reflect the Fock-state content of the projectile. The
non-perturbative information about the target wave function probed at a given
$Y$ is encoded in averages $\langle \ldots \rangle_Y$ taken with that target
wave function at $Y$ (denoted graphically by
$\diagram[height=.25cm]{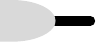}$ in the following). In the JIMWLK
formalism, this is phrased in terms of an average over the color source
distribution $Z_Y[U]$ (or equivalently $W_Y[\rho]$):
\begin{align}
  \label{eq:target-average}
  \langle \ldots \rangle_Y  = 
  \diagram[height=.3cm]{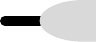} 
  \ldots 
  \diagram[height=.3cm]{targetwfshort-R}
= \int \Hat D[U]\ \ldots\ Z_Y[U] 
\ ,
\end{align}
where the dots ``$\ldots$'' stands for some generic $U-$content that depends
on the observable being considered.  What can be obtained by weak-coupling
methods within the JIMWLK formalism is the evolution of $\langle \ldots
\rangle_Y$ with the factorization rapidity $Y$.  The derivations of JIMWLK
evolution have mostly focused on the total cross-section, the literature on
evolution of diffractive observables in a JIMWLK context is very
terse~\cite{Hentschinski:2005er, Kovner:2006ge}. For this reason a detailed
explanation of how to extend JIMWLK evolution to observables beyond the total
cross-section is in order. Below, this will be done starting from an amplitude
or wave function picture for the total cross-section, in analogy
to~\cite{Kovner:2001vi}. In this framework it is easy to illustrate the
effects of restrictions on the final state (as constituted by the rapidity gap
events observed at HERA or the meson production cross-sections forming the
core of this exploration) on the structure of the evolution equation.

\begin{figure}[htb]
  \centering
\includegraphics{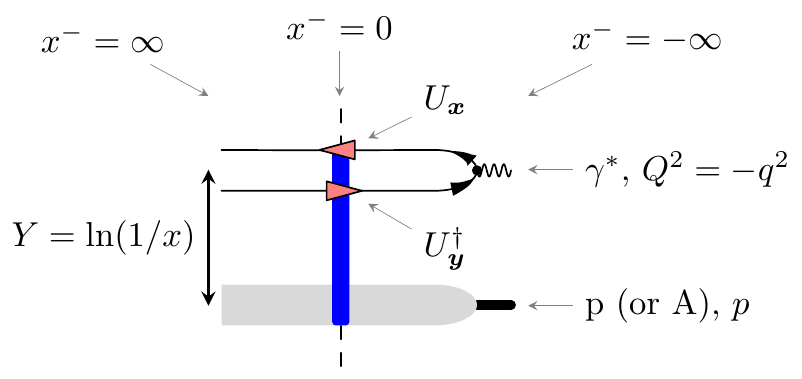}
\hfill
\includegraphics{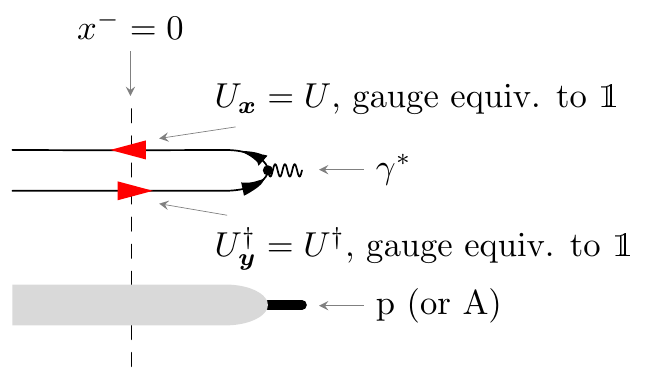}

\caption{Diagrammatic representation of the amplitude for $\gamma^* A$
  scattering at small $x$ at momentum transfer $Q^2=-q^2$. Light cone ``time''
  $x^-$ runs from right to left. The interacting ``out-state'' (left diagram)
  contains nontrivial interactions between projectile and target. The
  interaction region is indicated by a vertical bar (blue online) at $x^-=0$
  with superimposed explicit markers for the Wilson lines picked up by each
  projectile constituent. An arrow to the left indicates a $U$, an arrow to
  the right a $U^{-1}$. Arrows on gluon lines stand for Wilson lines in the
  adjoint representation.  The non-interacting ``in-state'' (right diagram)
  instead has no interactions and correspondingly constant Wilson line factors
  at $x^-=0$ which are gauge equivalent to the unit element.}
  \label{fig:gamma*-A-diagr-notation}
\end{figure}

Meson production cross-sections will be discussed below mainly in the case of
DIS, scattering a virtual photon $\gamma^*$ on a hadronic target with atomic
number $A$. Fig.~\ref{fig:gamma*-A-diagr-notation} provides the diagrammatic
notation used below to compactly summarize the contributions.

\subsection{Wilson line correlators in observables within and 
  beyond the total cross-section}
\label{sec:Wilson-lines-beyond-total-cross}

The small $x$ approximation to the DIS total cross-section at zeroth order in
$\ln(1/x)$ involves the eikonal scattering of the $q\Bar q$-Fock component of
the virtual photon on the target. The cross-section emerges from the absolute
value squared of the difference between the out-state (in which the $q\Bar q$
pair interacts with the target and picks up non-Abelian eikonal factors at
fixed transverse positions) and the in-state (where this interaction is
absent). Diagrammatically, one considers
\begin{subequations}
  \label{eq:zeroeth-order-cross}
\begin{equation}
  \label{eq:zeroeth-order-cross-a}
  \left\vert \diagram[width=1.2cm]{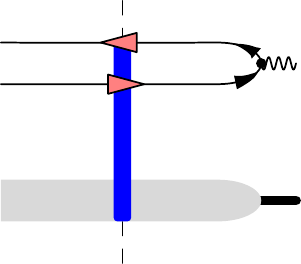}\right. - 
   \left. \diagram[width=1.2cm]{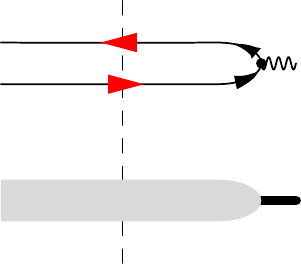} \right\vert^2
=\, 
\left( \diagram[width=1.2cm]{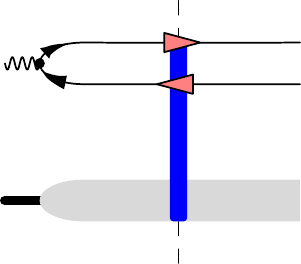} - \diagram[width=1.2cm]{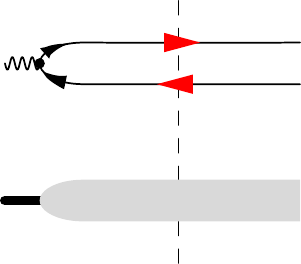} \right)
\left( \diagram[width=1.2cm]{qqbint-R} - \diagram[width=1.2cm]{qqbnoint-R}
\right)
\ .
\end{equation}
The unrestricted transverse momentum integrals will identify the transverse
coordinates left and right of the cut, so that the the $U$-content of these
diagrams is partially simplified:
\begin{equation}
  \label{eq:zeroeth-order-cross-b}
\eqref{eq:zeroeth-order-cross-a} =\, 
  \underset{\tr U_{\bm y}U^\dagger_{\bm x} U_{\bm x}U^\dagger_{\bm y}
    =\tr 1}{\diagram[width=1.2cm]{qqbint-L}
    \cut\diagram[width=1.2cm]{qqbint-R}}
  - \underset{\tr U_{\bm x}U^\dagger_{\bm y}}{
    \diagram[width=1.2cm]{qqbnoint-L}
    \cut
    \diagram[width=1.2cm]{qqbint-R}}
  - \underset{\tr U_{\bm y}U^\dagger_{\bm x}}{
    \diagram[width=1.2cm]{qqbint-L}
    \cut
    \diagram[width=1.2cm]{qqbnoint-R}}
  + \underset{\tr 1\vphantom{U^\dagger_{\bm y}}}{
    \diagram[width=1.2cm]{qqbnoint-L}
    \cut
    \diagram[width=1.2cm]{qqbnoint-R}}
\ .
\end{equation}
\end{subequations}
Note in particular that the $U$-factors left and right of the cut in the
$\overline{\text{out}}$-out overlap cancel against each
other so that
\begin{align}
  \label{eq:U-left-right-cancel}
  \diagram[width=1.2cm]{qqbint-L}
  \cut
  \diagram[width=1.2cm]{qqbint-R}
  = 
  \diagram[width=1.2cm]{qqbnoint-L}
  \cut
  \diagram[width=1.2cm]{qqbnoint-R}
\ ,
\end{align}
both contain no interaction with the target. The interaction with the target
is fully encoded in the $U$-content of the two remaining diagrams which takes
the form of two dipole operators
\begin{equation}
\label{eq:hatSqqbardef}
  \Hat S_{\bm x\bm y}^{q\Bar q} := \frac{\tr\left(U_{\bm x} U_{\bm
        y}^\dagger\right)}{N_c}
\hspace{1cm}\text{and}\hspace{1cm}
  \Hat S_{\bm y\bm x}^{q\Bar q} := \frac{\tr\left(U_{\bm y} U_{\bm
        x}^\dagger\right)}{N_c}
\ .
\end{equation}
In the expression for the total cross-section, these operators are averaged
over the target wave function, an operation that involves
both perturbative and non-perturbative information, that proves the most
difficult part of this calculation and induces the energy (or $Y$-) dependence
of the cross-section. The tool to extract this energy dependence is the JIMWLK
equation.

The total cross-section takes the form of a convolution between a wave
function overlap --involving the $q\Bar q$ component of the virtual photon
from right and left of the cut-- with the Wilson line correlators that
summarize the interaction with the target. The wave function overlap is the
same in all four terms of~(\ref{eq:zeroeth-order-cross}) and consist of both
transverse and longitudinal contributions.

In the present context the target averages of~(\ref{eq:hatSqqbardef}) satisfy
$\langle \Hat S_{\bm{x y}}^{q\Bar q} \rangle(Y) = \bigl[\langle \Hat S_{\bm{y
    x}}^{q\Bar q} \rangle(Y)\bigr]^*$, for all applications considered below
they will, in fact, be real. With this assumption, the four terms
in~(\ref{eq:zeroeth-order-cross-b}) assemble in such a way that the target
interaction of the total cross-section is fully summarized by the dipole
amplitude
\begin{align}
  \label{eq:dipole-amplitude}
  N_{\bm x \bm y, Y} :=   \frac{1}{N_c} 
  \left\langle\tr(1-U_{\bm{x}} U^\dagger_{\bm{y}})
  \right\rangle_Y
\ ,
\end{align}
and one has re-derived the general formula for the total cross-section at
small $x$,
\begin{equation}
  \label{eq:dipole-cross}
  \sigma^{\mathrm{DIS}}(Y,Q^2) 
          =2\int\!\!d^2 r
  \int\limits^1_0\!d\alpha\
|\Psi(\alpha,r^2,Q^2)|^2
  \int d^2b \ N_{\bm x\bm y, Y}
\ ,
\end{equation}
where $\bm{r}=\bm{x}-\bm{y}$ and $\bm b =(\bm x+\bm y)/2$ denote dipole size
and impact parameter. At fixed $Q^2$,
$|\Psi(\alpha,r^2,Q^2)|^2$ encodes the probability to find a $q\Bar q$ pair of
size $|\bm{r}|$ and longitudinal momentum fraction $\alpha$ inside the virtual
photon.  The impact-parameter-integrated dipole amplitude has the
interpretation of a $q\Bar q$-dipole cross-section on the target.

JIMWLK-evolution has been originally derived for the total cross-section via
the optical theorem~\cite{Weigert:2000gi,Balitsky:1996ub} and applies directly
to the evolution of the total cross-section in
Eq.~\eqref{eq:zeroeth-order-cross} or~\eqref{eq:dipole-cross}.

The derivation of the expressions for the zeroth order cross-section used here
avoids using the optical theorem. This allows for easier generalization to
other, more differential observables. Such observables can also be addressed
using JIMWLK evolution, their treatment, however can only be found either in
very compact form~\cite{Hentschinski:2005er}, or in a somewhat abstract
formalism~\cite{Kovner:2006ge} -- the treatment here will address all these
quantities in a simple unified formalism. The re-derivation of the evolution
equations will be done at the one loop level only, with the main goal of
identifying the underlying structures and requirements that will have to be
met at any loop order.

As already mentioned in the introduction, inclusive vector-meson production is
an example for such more differential observables and is of particular
phenomenological interest.  Addressing inclusive vector-meson production
cross-sections requires only a formally trivial modification of the expression
in Eq.~\eqref{eq:zeroeth-order-cross}: one only needs to project the $q\Bar q$
states in the final state onto the vector-meson state under consideration,
without imposing any further restrictions on the final state. This amounts to
replacing~(\ref{eq:zeroeth-order-cross}) by
\begin{align}
  \label{eq:vm-inclusive-zeroeth}
  \Biggl[
  \diagram[height=1.1cm]{qqbint-L}-\diagram[height=1.1cm]{qqbnoint-L}
  \Biggr] 
  \diagram[height=1.1cm]{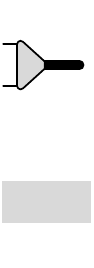}
  \cut
  \diagram[height=1.1cm]{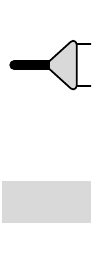}
  \Biggl[
  \diagram[height=1.1cm]{qqbint-R}-\diagram[height=1.1cm]{qqbnoint-R}
  \Biggr]
 \ .
\end{align}
Expanding the product, one finds that only the operator content of the
$\overline{\text{out}}$-out overlap is changed markedly from its counterpart
in the total cross-section.  Since the $q\Bar q$ in the final state are now
projected onto the vector-meson wave function (instead of being integrated over
all transverse momenta) the coordinates of the Wilson lines right and left of
the cut will not become equal. Also the sum over all possible colors is
replaced by a projection onto a singlet combination, effectively separating
the single color trace appearing in the first term of
Eq.~(\ref{eq:zeroeth-order-cross-b}) into a product of traces.

Since there is no further restriction of the final state, these two traces
will remain in a common average, as induced by the single pair of target
states present in Eq.~(\ref{eq:vm-inclusive-zeroeth}). This leaves us with a
nontrivial correlator in the $\overline{\text{out}}$-out overlap, namely
\begin{equation}
  \label{eq:qqb2-corr-gen}
  \langle
  \frac{\tr(U_{\bm y'}U^\dagger_{\bm x'})}{N_c} 
  \frac{\tr(U_{\bm x}U^\dagger_{\bm y})}{N_c} 
  \rangle_Y 
  = 
  \langle \Hat S_{\bm y'\bm x'}^{q\Bar q}\Hat S_{\bm x\bm y}^{q\Bar q}\rangle_Y
\ .
\end{equation}
For consistency of the JIMWLK-formalism, one would expect that both the two
\emph{and} four point correlators entering~\eqref{eq:vm-inclusive-zeroeth}
evolve under JIMWLK evolution. This is indeed generally assumed in the
literature. Since the Wilson line factors originate from both sides of the
cut, this is not entirely obvious. Sec.~\ref{sec:mueller-optical} takes a
closer look at how JIMWLK-evolution emerges in the present S-matrix formalism
(without recourse to the optical theorem), first for the total cross-section
and then for the non-trivial evolution of the \tol{out}-out-overlap in
Eq.~\eqref{eq:vm-inclusive-zeroeth}. The argument employs
``flip-back-identities'' sometimes  referred to as the Mueller
optical theorem to confirm that~\eqref{eq:qqb2-corr-gen} indeed evolves as
four Wilson-line operator under JIMWLK-evolution.

The argument also clarifies that the correlators in the remaining
contributions, \tol{in}-in-, \tol{in}-out-, and \tol{out}-in-overlaps and
their energy dependence do not change compared to their total cross-section
counterpart.

Note that the large-$N_c$ limit (the BK-approximation) sidesteps all these
issues by factoring the correlator:
\begin{align}
  \label{eq:qqb2-corr-gen-Nc-exp}
  \langle
  \frac{\tr(U_{\bm y'}U^\dagger_{\bm x'})}{N_c} 
  \frac{\tr(U_{\bm x}U^\dagger_{\bm y})}{N_c} 
  \rangle_Y
  & = \langle
  \frac{\tr(U_{\bm y'}U^\dagger_{\bm x'})}{N_c}
  \rangle_Y \langle
  \frac{\tr(U_{\bm x}U^\dagger_{\bm y})}{N_c} 
  \rangle_Y
  +{\cal O}(1/N_c)
\notag \\ &
 =
  \langle \Hat S_{\bm y'\bm x'}^{q\Bar q}\rangle_Y \langle 
  \Hat S_{\bm x\bm y}^{q\Bar q}\rangle_Y+{\cal O}(1/N_c)
\ .
\end{align}

In calculating the analytical form for the exclusive vector-meson from the
diagrams in Eq.~\eqref{eq:vm-inclusive-zeroeth} one encounters wave function
factors of the virtual photons in the initial states as well as those of the
vector-meson in the final states in each of the amplitudes left and right of
the cut.  All four terms in~\eqref{eq:vm-inclusive-zeroeth} are convoluted
with \emph{two separate} wave function overlap factors, originating from the
right and left side of the cut respectively. They depend on the photon
polarization, quark masses, phenomenological mass and size parameters of the
mesons and dynamical variables, longitudinal momentum fraction $\alpha$,
inter-quark-distance $\bm r$, and momentum transfer $Q^2$ and will be denoted
$\Psi_T(\alpha,\bm r,Q^2)$ and $\Psi_L(\alpha,\bm r,Q^2)$
respectively. Explicit expressions are given in
App.~\ref{sec:photon-vector-meson-wave-function-overlaps}.

With the convention $t=-\bm l^2$, the expression for the inclusive
vector-meson production cross-section takes the form~\cite{Marquet:2009vs}
\begin{align}
\label{eq:vm-inclusive-zeroth-analytical}
  4\pi \frac{d\sigma_{T,L}}{d t} = &
  \int\limits_0^1 d\alpha d\alpha' \int d^2x d^2x' d^2y d^2 y'
  e^{-i \bm l\cdot[
    (\alpha\bm x+(1-\alpha)\bm y)
    -(\alpha'\bm x'+(1-\alpha')\bm y')
    ]
  }
\notag \\ & {}\times
  \Psi_{T,L}^*(\alpha',\bm x'-\bm y',Q^2)
  \Psi_{T,L}(\alpha,\bm x-\bm y,Q^2)
  \notag \\ & {}\times
  \biggl[\langle  \tr(U_{\bm y'} U_{\bm x'}^\dagger) 
                          \tr(U_{\bm x} U_{\bm y}^\dagger)
         \rangle_Y/N_c^2
         - \langle \tr(U_{\bm y'} U_{\bm x'}^\dagger) \rangle_Y/N_c
         -
         \langle
                 \tr(U_{\bm x} U_{\bm y}^\dagger)\rangle_Y/N_c
         +1
  \biggr]
\ .
\end{align}

The cross-section for \emph{exclusive} vector-meson production, in which one
requires the target to remain intact, on the other hand, contains two separate
$U$-field averages on the level of the amplitudes without any need for an
additional approximation (such as Eq.~(\ref{eq:qqb2-corr-gen-Nc-exp})).
Eq.~\eqref{eq:vm-inclusive-zeroeth} is replaced by
\begin{align}
  \label{eq:vm-exclusive-zeroeth}
  \Biggl[
  \diagram[height=1.1cm]{qqbint-L}-\diagram[height=1.1cm]{qqbnoint-L}
  \Biggr] 
  \diagram[height=1.1cm]{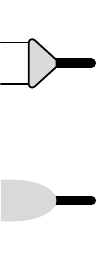}
  \cut
  \diagram[height=1.1cm]{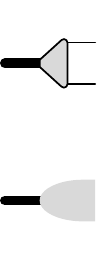}
  \Biggl[
  \diagram[height=1.1cm]{qqbint-R}-\diagram[height=1.1cm]{qqbnoint-R}
  \Biggr]
\end{align}
and differs from~\eqref{eq:vm-inclusive-zeroeth}
(and~\eqref{eq:vm-inclusive-zeroth-analytical}) only in the $U$-field average
of the \tol{out}-out-overlap  which now is factorized into
\begin{equation}
  \label{eq:qqb2-corr-gen-fact}
  \langle
  \frac{\tr(U_{\bm y'}U^\dagger_{\bm x'})}{N_c} 
  \rangle_Y
  \langle
  \frac{\tr(U_{\bm x}U^\dagger_{\bm y})}{N_c} 
  \rangle_Y 
  = 
  \langle \Hat S_{\bm y'\bm x'}^{q\Bar q}\rangle_Y 
  \langle\Hat S_{\bm x\bm y}^{q\Bar q}\rangle_Y
\ .
\end{equation}
The arguments given in Sec.~\ref{sec:mueller-optical} ensure that these indeed
evolve independently under JIMWLK-evolution as one would expect. Due to this
factorization of correlators, also the full analytic expression corresponding
to~\eqref{eq:vm-exclusive-zeroeth} factorizes into independent complex
conjugate amplitude-factors:
\begin{align}
\label{eq:vm-exclusive-zeroth-analytical}
  4\pi \frac{d\sigma_{T,L}}{d t} = \left|
  \int\limits_0^1 d\alpha \int d^2x  d^2y\
  e^{-i \bm l\cdot[
    (\alpha\bm x+(1-\alpha)\bm y)
    ]
  }\ \Psi_{T,L}(\alpha,\bm x-\bm y,Q^2)\ N_{\bm x \bm y, Y} 
  \right|^2
\ .
\end{align}
Note that by taking the difference between~\eqref{eq:vm-inclusive-zeroeth}
and~\eqref{eq:vm-exclusive-zeroeth} one directly \emph{measures} correlator
factorization violations of the form
\begin{align}
  \label{eq:corr-fact-viol}
  \langle  \tr(U_{\bm y'} U_{\bm x'}^\dagger) 
                   \tr(U_{\bm x} U_{\bm y}^\dagger)
         \rangle_Y/N_c^2
  -\langle  \tr(U_{\bm y'} U_{\bm x'}^\dagger) 
         \rangle_Y \langle
                 \tr(U_{\bm x} U_{\bm y}^\dagger)
         \rangle_Y/N_c^2
\ ,
\end{align}
a special case of which [$\bm x'=\bm x$] have been extensively discussed
in~\cite{Kovchegov:2008mk} from a theoretical perspective in the light of full
JIMWLK evolution.  Measuring such differences would allow a direct
\emph{experimental} test of rather subtle features of JIMWLK evolution. These
features are of particular interest in that they evidently lie beyond the
BK-approximation, for
which~(\ref{eq:corr-fact-viol})vanishes identically.

Inclusive and exclusive vector-meson production are by no means the only
observables that can be addressed in this way. For example heavy meson 
production cross-sections are closely related as far as the Wilson line
correlators are concerned, despite the change in projectile required. For the
inclusive case one would consider
\begin{align}
  \label{eq:heavy-inclusive-zeroeth}
  \overset{in}{
  \diagram[height=1.1cm]{vmwf-t-L}
  }
  \Biggl[
  \diagram[height=1.1cm]{qqbintreg-L}-\diagram[height=1.1cm]{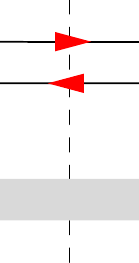}
  \Biggr] 
  \overset{Q,\Bar q}{
  \diagram[height=1.1cm]{vmwf-R}
  }
  \cut
  \overset{Q,\Bar q}{
  \diagram[height=1.1cm]{vmwf-L}
  }
  \Biggl[
  \diagram[height=1.1cm]{qqbintreg-R}-\diagram[height=1.1cm]{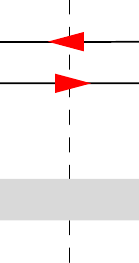}
  \Biggr]
  \overset{in}{
  \diagram[height=1.1cm]{vmwf-t-R}
  }
\ ,
\end{align}
where the state marked ``\emph{in}'' may be some hadron with quantum numbers
suitable to produce a meson with a heavy quark $Q$. In the phase space region
where the mass $m_Q$ can be considered large, the transverse position of the
corresponding Wilson lines left and right of the cut will be approximately the
same, one encounters the coincidence-limit of~\eqref{eq:qqb2-corr-gen} at $\bm
x' = \bm x$ which inherits its energy dependence from the more general case,
but takes a simpler form.

\subsection{Evolution and the Mueller optical theorem}
\label{sec:mueller-optical}

The main subtlety with generalizing JIMWLK evolution for the total
cross-section (as initially derived using the optical theorem) to the case of
more differential observables is the fact that this derivation builds on a
whole set of real-virtual cancellations automatically taken care of by the
optical theorem, which for more differential observables are no longer valid.
Already for a description of rapidity gap events at HERA, the restriction of
the final state leads to a modification of the overall evolution equation of
the diffractive cross-section that at first sight appears to be only
indirectly related to JIMWLK (or BK) evolution [see for
example~\cite{Kovchegov:1999ji}, Eq.~(11)]. Nevertheless it can be decomposed
into individual contributions akin to the terms
of~(\ref{eq:vm-inclusive-zeroeth}) that still follow JIMWLK (or BK) evolution,
subject to specific initial conditions~\cite{Kovchegov:1999ji,
  Hentschinski:2005er}.

\begin{figure}[htb]
  \centering
\includegraphics{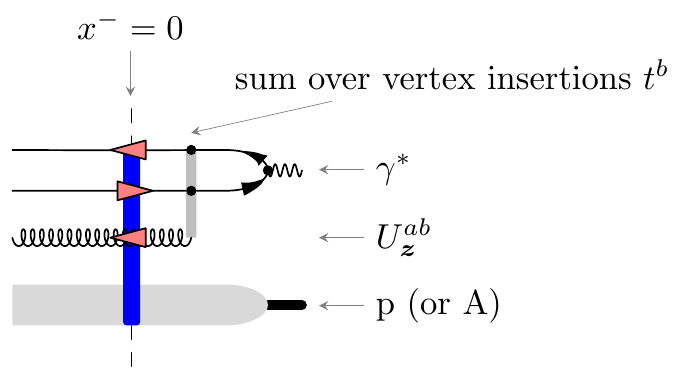}

\caption{Diagrammatic notation for gluon emission at small $x$. Gluon lines
  carry an adjoint Wilson line $U_{\bm z}^{a b}$, the vertical gray lines
  indicate sums over vertex insertions at the black vertex dots. 
    }
  \label{fig:qqb+g-diagram-notation}
\end{figure}

To address JIMWLK-corrections to the observables in
Sec.~\ref{sec:Wilson-lines-beyond-total-cross} one needs to include gluon
emission from the zeroth order expressions shown
there. Fig.~\ref{fig:qqb+g-diagram-notation} shows the diagrammatic shorthand
notations employed to that end in an exemplary fashion.

\begin{figure}[htb]
  \centering
\includegraphics{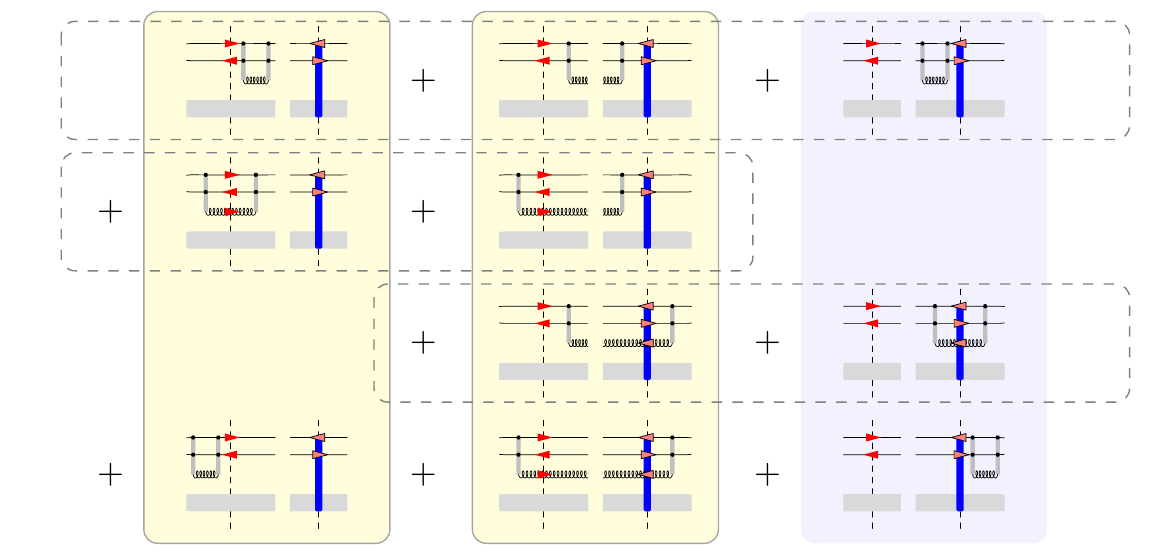}
  
  \caption{Cancellation patterns in the \tol{in}-out-overlap: 
    Diagrams in the first two columns (full outlines) 
    cancel against each other according to
    Eq.~\eqref{eq:x^-cancellations}. This guarantees that the JIMWLK
    Hamiltonian yields zero when acting on constant $U$-fields.
    Alternatively the diagrams in the first three lines cancel amongst each
    other in a prototypical final state cancellation (dashed outlines).
    The remaining diagrams in both cases reflect the leading order JIMWLK
    corrections to the total
    DIS cross-section.
  }
  \label{fig:inbar-out-cancellation-patterns}
\end{figure}

There is a strong set of regularities in the small $x$ corrections that is
intimately connected with the properties of the JIMWLK-Hamiltonian.  The full
pattern is already visible if one considers the leading order (LO) corrections
to the $\overline{\text{in}}$-out overlap in
Eq.~(\ref{eq:zeroeth-order-cross}), which consists of the diagrams shown in
Fig.~\ref{fig:inbar-out-cancellation-patterns} (with the photon- and
target-wave-function factored -- the relationships remain valid even if the
$q\Bar q$ is in an octet state)

In Fig.~\ref{fig:inbar-out-cancellation-patterns},
the first two columns cancel amongst themselves. Technically, this is due to
the diagrammatic identities shown in detail in the left column of
Eq.~(\ref{eq:x^-cancellations})
\begin{subequations}
    \label{eq:x^-cancellations}
\begin{align}
  \label{eq:x^-cancellations-a}
 \diagram[height=1.1cm]{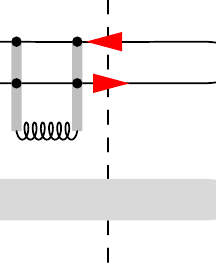} 
& = 
  \diagram[height=1.1cm]{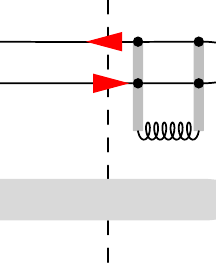}
= 
 -\frac12\diagram[height=1.1cm]{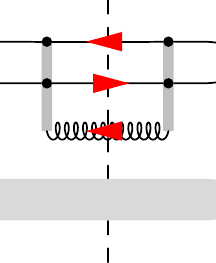} 
& \Rightarrow  \hspace{1cm} &
\diagram[height=1.1cm]{qqbnointreg-rr-R} 
+ 
\diagram[height=1.1cm]{qqbnointreg-lr-R} 
+
\diagram[height=1.1cm]{qqbnointreg-ll-R} = 0
\ ,
\\
\label{eq:x^-cancellations-b}
 \diagram[height=1.1cm]{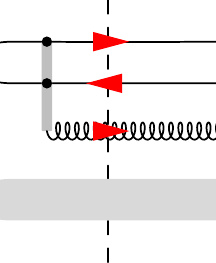}
& = -
  \diagram[height=1.1cm]{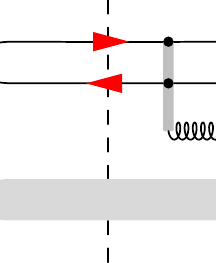}
& 
\Rightarrow \hspace{1cm} &
\diagram[height=1.1cm]{qqbnointreg-lf-L}
+
\diagram[height=1.1cm]{qqbnointreg-rf-L} = 0
\ .
\end{align}
\end{subequations}
Eq.~(\ref{eq:x^-cancellations-b}) states that there without interaction with
the target, no gluons are emitted into the final
state. Eq.~(\ref{eq:x^-cancellations-a}), on the other hand, expresses the
fact that there is no JIMWLK evolution in the absence of interaction with the
target -- for constant $U$-fields which are gauge equivalent to $U=1$.  This
becomes obvious once one recognizes that the remaining third column contains
the JIMWLK contributions to the evolution of $S^{q\Bar q}$ as originally
calculated via the optical theorem in~\cite{Weigert:2000gi,Balitsky:1996ub}:
\begin{equation}
  \label{eq:H_JIMWLK-barin-out-alt}
\ln(1/x)  H_{\text{JIMWLK}} 
\diagram[height=1.1cm]{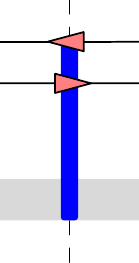}
=
\diagram[height=1.1cm]{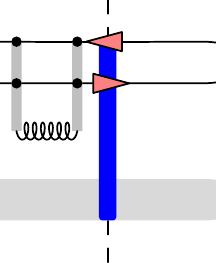}
+
\diagram[height=1.1cm]{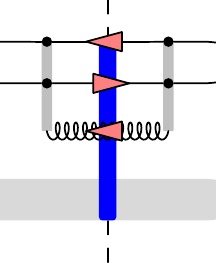}
+
\diagram[height=1.1cm]{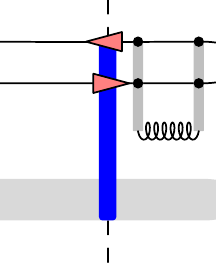}
\ .
\end{equation}
Alternatively, one may notice that the diagrams of each of the first three
lines in Fig.~\ref{fig:inbar-out-cancellation-patterns} separately sum to zero
following the prototypical pattern of final state cancellations. This recasts
the contributions to JIMWLK evolution in terms of the diagrams of the last
line in Fig.~\ref{fig:inbar-out-cancellation-patterns} and offers an
alternative interpretation with gluons emitted into the final state during
evolution.

The same argument trivially ensures that all corrections to the
\tol{in}-in-overlap must vanish as well.

For the total cross-section, in fact, only the mixed \tol{in}-out- and
\tol{out}-in-overlaps receive small-$x$ corrections, also the corrections to
the \tol{out}-out-overlap vanish altogether: Arranging the contributions as in
Fig.~\ref{fig:inbar-out-cancellation-patterns}, one first notices that there
are no cancellations of columns -- interaction with the target creates
JIMWLK-contributions in the first and last column, and also allows a gluon to
appear in the final state of the middle column.  However, the first three
lines cancel in the sense of final state cancellations just as for the
\tol{in}-out-overlap. In addition, all $U$ fields in the last line cancel (c.f
Eq.~\eqref{eq:U-left-right-cancel}), mapping the contribution onto a
contribution equivalent to the JIMWLK-diagrams with constant $U$-fields,
setting the complete set of corrections to the \tol{out}-out-overlap to
zero. 

One recovers the established result for JIMWLK evolution of the total
$\gamma^* A$-cross-section:
\begin{align}
  \label{eq:JIMWLK-via-diagrams}
  \frac{d}{dY} 
\diagram[height=1.1cm]{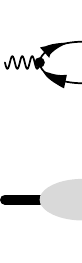}
\left[\diagram[height=1.1cm]{qqbintreg-L} 
- 
\diagram[height=1.1cm]{qqbnointreg-L} 
\right]
\left[
\diagram[height=1.1cm]{qqbintreg-R} 
- 
\diagram[height=1.1cm]{qqbnointreg-R} 
\right]
\diagram[height=1.1cm]{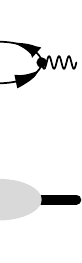}
= 
\diagram[height=1.1cm]{qqbt-wf-L}
H_{\text{JIMWLK}}
\left[\diagram[height=1.1cm]{qqbintreg-L} 
- 
\diagram[height=1.1cm]{qqbnointreg-L} 
\right]
\left[
\diagram[height=1.1cm]{qqbintreg-R} 
- 
\diagram[height=1.1cm]{qqbnointreg-R} 
\right]
\diagram[height=1.1cm]{qqbt-wf-R}
\ .
\end{align}

If one considers the corrections to the vector-meson production cross-sections
of Eq.~\eqref{eq:vm-inclusive-zeroeth} and~\eqref{eq:vm-exclusive-zeroeth},
the argument for the mixed \tol{in}-out- and \tol{out}-in-overlaps still
follow the pattern of Fig.~\ref{fig:inbar-out-cancellation-patterns} since
they are driven by cancellations based on~\eqref{eq:x^-cancellations} that
occur within one non-interacting amplitude factor. What changes are the
corrections to the \tol{out}-out-overlaps to be discussed next.

The cancellation of $U$-factors left and right of the cut, that occur in the
corresponding contribution to the total cross-section, no longer takes place:
Both, the transverse positions of the $U$-fields left and right of the
cut differ from each other and the color indices are contracted, not into each
other, but into the meson wave function in the final state.  Instead,
corrections to the \tol{out}-out overlap of the inclusive vector-meson
cross-section in fact evolve according to JIMWLK-evolution of a four
Wilson-line correlator, as given by
\begin{align}
  \label{eq:qqbqqb-amp-JIMWLK}
  \ln(1/x) H_{\text{JIMWLK}} 
\diagram[height=1.3cm]{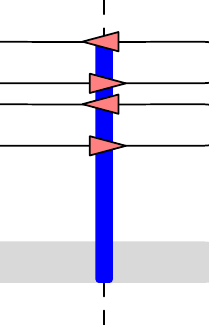} 
=
 \diagram[height=1.3cm]{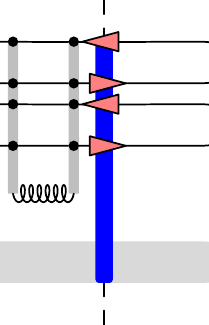}
+
 \diagram[height=1.3cm]{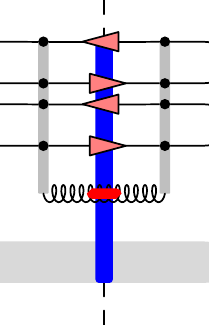}
+
 \diagram[height=1.3cm]{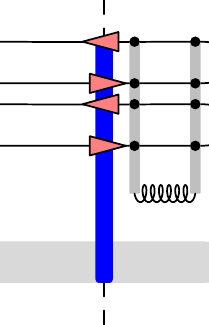}
\end{align} 
since the contributions on the right hand side of
Eq.~\eqref{eq:qqbqqb-amp-JIMWLK} unfold precisely into the expected list of
diagrams:\footnote{ The numbers below the diagrams stand for (symmetry
  factors)$\cdot$(number of terms from vertex sums) to help relate the
  contributions. Note in particular that the gluon Wilson-lines in the top
  middle diagram cancel between the amplitude and its complex conjugate since
  the gluon's final state phase space remains unrestricted.}
\begin{subequations}
  \label{eq:inclusive-vm-flipback-corresp}
\begin{align}
 \underset{16}{
    \diagram[height=1.3cm]{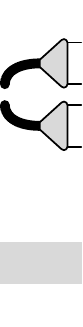}
     \diagram[height=1.3cm]{qqbqqbintreg-ll-R}
 }
 = & \underset{4}{
  \diagram[height=1.1cm]{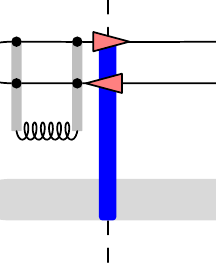}
   \diagram[height=1.1cm]{vmwf-R}
  \,
  \diagram[height=1.1cm]{vmwf-L}
  \diagram[height=1.1cm]{qqbintreg-R}} 
 + \underset{2\cdot 4}{
    \diagram[height=1.1cm]{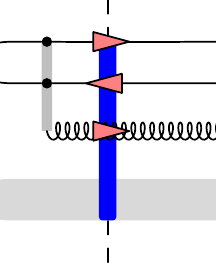}
     \diagram[height=1.1cm]{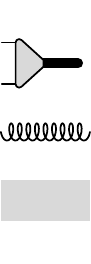}
  \,
  \diagram[height=1.1cm]{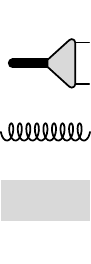}\diagram[height=1.1cm]{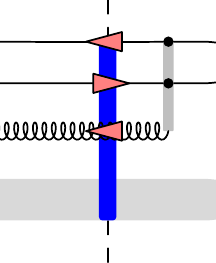}} 
 + \underset{4}{\diagram[height=1.1cm]{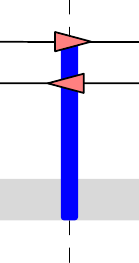}
     \diagram[height=1.1cm]{vmwf-R}
  \,
  \diagram[height=1.1cm]{vmwf-L}\diagram[height=1.1cm]{qqbintreg-ll-R}} 
\\
 \underset{2\cdot 16}{
    \diagram[height=1.3cm]{vmwf2-t-L}
    \diagram[height=1.3cm]{qqbqqbintreg-lr-R}
 }  
 = & \underset{2\cdot 4}{\diagram[height=1.1cm]{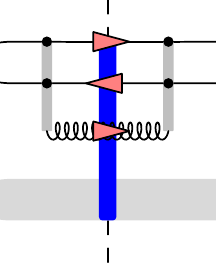}
   \diagram[height=1.1cm]{vmwf-R}
\,
  \diagram[height=1.1cm]{vmwf-L}\diagram[height=1.1cm]{qqbintreg-R}} 
+\underset{2\cdot 4}{\diagram[height=1.1cm]{qqbintreg-lf-L}
   \diagram[height=1.1cm]{vmwf+g-R}
\,
  \diagram[height=1.1cm]{vmwf+g-L}
  \diagram[height=1.1cm]{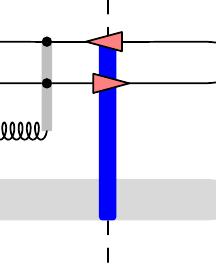}} 
\notag
\\
& \phantom{ \underset{4}{
  \diagram[height=1.1cm]{qqbintreg-ll-L}
   \diagram[height=1.1cm]{vmwf-R}
  \,
  \diagram[height=1.1cm]{vmwf-L}
  \diagram[height=1.1cm]{qqbintreg-R}} }
 +\underset{2\cdot 4}{\diagram[height=1.1cm]{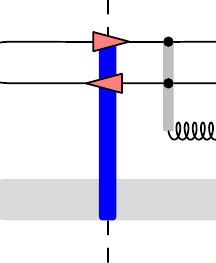}
   \diagram[height=1.1cm]{vmwf+g-R}
\,
  \diagram[height=1.1cm]{vmwf+g-L}\diagram[height=1.1cm]{qqbintreg-lf-R}} 
 +\underset{2\cdot 4}{\diagram[height=1.1cm]{qqbintreg-L}
   \diagram[height=1.1cm]{vmwf-R}
\,
  \diagram[height=1.1cm]{vmwf-L}\diagram[height=1.1cm]{qqbintreg-lr-R}} 
\\
 \underset{16}{
   \diagram[height=1.3cm]{vmwf2-t-L}
   \diagram[height=1.3cm]{qqbqqbintreg-rr-R}
 } 
 = &
\underset{4}{\diagram[height=1.1cm]{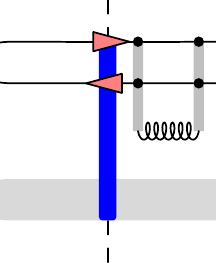}
   \diagram[height=1.1cm]{vmwf-R}
\,
  \diagram[height=1.1cm]{vmwf-L}\diagram[height=1.1cm]{qqbintreg-R}} 
 +\underset{2\cdot 4}{\diagram[height=1.1cm]{qqbintreg-rf-L}
   \diagram[height=1.1cm]{vmwf+g-R}
\,
  \diagram[height=1.1cm]{vmwf+g-L}\diagram[height=1.1cm]{qqbintreg-rf-R}}  
 +\underset{4}{\diagram[height=1.1cm]{qqbintreg-L}
   \diagram[height=1.1cm]{vmwf-R}
\,
  \diagram[height=1.1cm]{vmwf-L}\diagram[height=1.1cm]{qqbintreg-rr-R}} 
\end{align}
\end{subequations}
The diagrammatic identification is straightforward and holds for specific
pairs of diagrams in the sums over insertions displayed on both the left and
right hand side of the equations in~\eqref{eq:inclusive-vm-flipback-corresp}.
Eq.~(\ref{eq:inclusive-vm-flipback-corresp}) lists only one loop correlators,
but the situation is prototypical and generalizes to arbitrary orders.

One may dub relations of the type shown in
Eq.~\eqref{eq:inclusive-vm-flipback-corresp} ``flip-back-identities'' as they
relate correlators with (anti-) quark Wilson-lines on both sides of the final
state cut (on the right), to equivalent ones that can be thought of as having
those lines only on one side of the cut. Note, however, that in doing so one
reinterprets a quark Wilson line $U_{\bm x}^\dagger$ in the complex conjugate
amplitude as an anti-quark line in an amplitude and vice versa.

Also the term Mueller optical theorem has been used to refer to this type of
relationship. Its diagrammatic content is by no means trivial: It is
remarkable that diagrams like those in the middle column of the right hand
side, with an additional gluon in the final state, map onto amplitudes in
which no gluon is emitted.

The contributions sum into
\begin{align}
  \label{eq:vm-inclusive-evo}
  \frac{d}{d Y} &
  \diagram[height=1.1cm]{qqbt-wf-L} 
  \Biggl[
  \diagram[height=1.1cm]{qqbintreg-L}-\diagram[height=1.1cm]{qqbnointreg-L}
  \Biggr] 
  \diagram[height=1.1cm]{vmwf-R}
  \cut
  \diagram[height=1.1cm]{vmwf-L}
  \Biggl[
  \diagram[height=1.1cm]{qqbintreg-R}-\diagram[height=1.1cm]{qqbnointreg-R}
  \Biggr]
  \diagram[height=1.1cm]{qqbt-wf-R} 
\notag \\ &
   =\diagram[height=1.1cm]{qqbt-wf-L} 
\Biggl\{
H_{\text{JIMWLK}}
   \Biggl[
  \diagram[height=1.1cm]{qqbintreg-L}-\diagram[height=1.1cm]{qqbnointreg-L}
  \Biggr] 
  \diagram[height=1.1cm]{vmwf-R}
  \cut
  \diagram[height=1.1cm]{vmwf-L}
  \Biggl[
  \diagram[height=1.1cm]{qqbintreg-R}-\diagram[height=1.1cm]{qqbnointreg-R}
  \Biggr]
\Biggr\}
  \diagram[height=1.1cm]{qqbt-wf-R} 
\ ,  
\end{align}
which, as outlined in Eqns.~\eqref{eq:qqbqqb-amp-JIMWLK}
and~\eqref{eq:inclusive-vm-flipback-corresp} contains JIMWLK-evolution of
four-Wilson-line operators in the \tol{out}-out-overlaps.

\begin{figure}[htb]
  \centering

\includegraphics{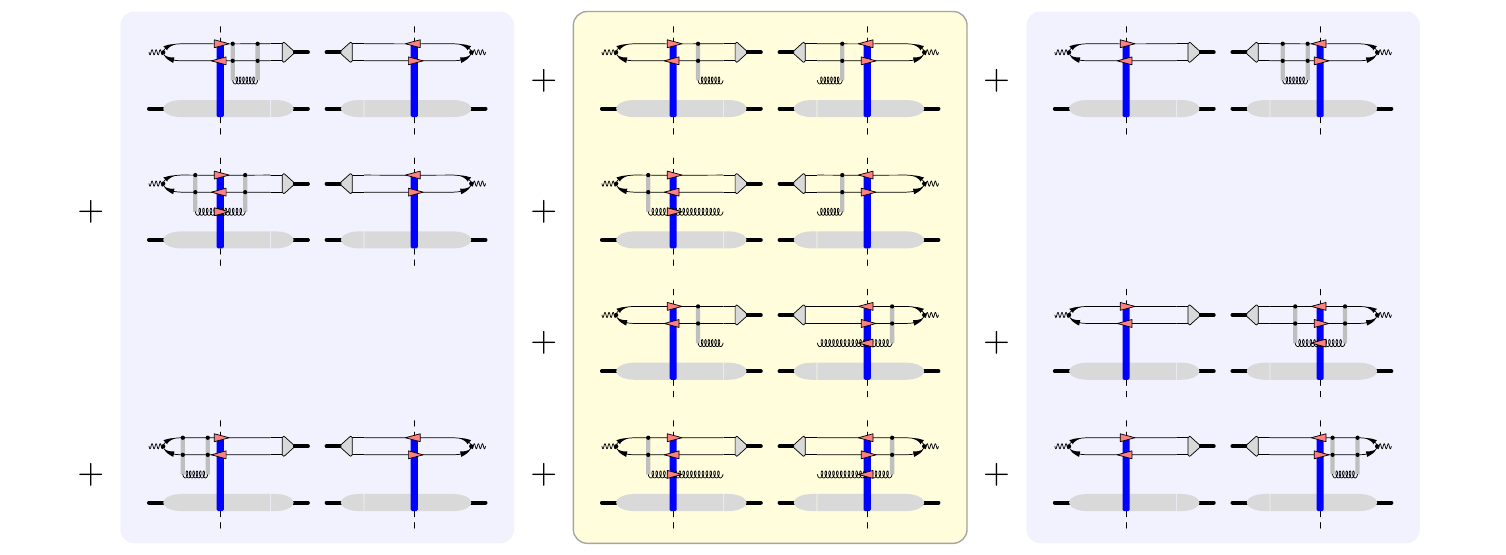}  

\caption{The corrections to the \tol{out}-out-overlap for exclusive
  vector-meson production. With both the initial- and final-states of
  projectile and target in both amplitudes projected onto color singlets, the
  final state emissions in the middle column are forced to zero by color
  conservation.}
  \label{fig:outbar-out-exclusive-vector-meson-production}
\end{figure}

This is notably different from the corrections to the \tol{out}-out-overlap
for exclusive vector-meson production in which all emission of gluons into the
final state are prohibited by color conservation as shown in
Fig.~\ref{fig:outbar-out-exclusive-vector-meson-production}. Contrary to
Eqns.~\eqref{eq:inclusive-vm-flipback-corresp},~\eqref{eq:qqbqqb-amp-JIMWLK},
and~\eqref{eq:H_JIMWLK-barin-out-alt}, this argument requires the projection
onto separate singlets of projectile and target in both the initial- and
final-states.
Since emission into the final state is prohibited, evolution of the
\tol{out}-out-overlap factorizes into separate evolution for the amplitude and
its complex conjugate, as anticipated already upon examining the zeroth order
correlator. The corresponding evolution equation is given by
\begin{align}
  \label{eq:vm-exclusive-evo}
  \frac{d}{d Y} &
  \diagram[height=1.1cm]{qqbt-wf-L} 
  \Biggl[
  \diagram[height=1.1cm]{qqbintreg-L}-\diagram[height=1.1cm]{qqbnointreg-L}
  \Biggr] 
  \diagram[height=1.1cm]{vmwf-t-R}
  \cut
  \diagram[height=1.1cm]{vmwf-t-L}
  \Biggl[
  \diagram[height=1.1cm]{qqbintreg-R}-\diagram[height=1.1cm]{qqbnointreg-R}
  \Biggr]
  \diagram[height=1.1cm]{qqbt-wf-R} 
\notag \\ &
   =\diagram[height=1.1cm]{qqbt-wf-L} 
\Biggl\{
H_{\text{JIMWLK}}
   \Biggl[
  \diagram[height=1.1cm]{qqbintreg-L}-\diagram[height=1.1cm]{qqbnointreg-L}
  \Biggr]
\Biggr\} 
  \diagram[height=1.1cm]{vmwf-t-R}
  \cut
  \diagram[height=1.1cm]{vmwf-t-L}
  \Biggl[
  \diagram[height=1.1cm]{qqbintreg-R}-\diagram[height=1.1cm]{qqbnointreg-R}
  \Biggr]
  \diagram[height=1.1cm]{qqbt-wf-R} 
\notag \\ &
+\diagram[height=1.1cm]{qqbt-wf-L} 
   \Biggl[
  \diagram[height=1.1cm]{qqbintreg-L}-\diagram[height=1.1cm]{qqbnointreg-L}
  \Biggr] 
  \diagram[height=1.1cm]{vmwf-t-R}
  \cut
  \diagram[height=1.1cm]{vmwf-t-L}
\Biggl\{
H_{\text{JIMWLK}}
  \Biggl[
  \diagram[height=1.1cm]{qqbintreg-R}-\diagram[height=1.1cm]{qqbnointreg-R}
  \Biggr]
\Biggr\}
  \diagram[height=1.1cm]{qqbt-wf-R} 
\ .  
\end{align}
Eq.~\eqref{eq:vm-exclusive-evo} states that the factorized correlators
entering~\eqref{eq:vm-exclusive-zeroeth}
and~\eqref{eq:vm-exclusive-zeroth-analytical} indeed evolve independently with
JIMWLK. The proof given in
Fig.~\ref{fig:outbar-out-exclusive-vector-meson-production} is strictly one
loop and needs to be reexamined already at two loop order, where two gluons
can enter the final state and potentially form a color singlet.  To perform
this analysis efficiently, it becomes essential to translate the NLO results of
Balitsky and Chirilli~\cite{Balitsky:2008zz, Balitsky:2008rc, Balitsky:2009xg,
  Balitsky:2009yp} into JIMWLK language -- a task well beyond the scope of
questions addressed here.

\section{The Gaussian truncation of JIMWLK evolution}
\label{sec:corr-gauss-trunc}

\subsection{Evolution of correlators}
\label{sec:evol-corr}

Solving the JIMWLK equations to obtain solutions for the vector-meson
production observables introduced above is at present curtailed by severe
practical limitations -- truncations of JIMWLK-evolution, however, can be
successfully confronted with data.  Aside from the BK-approximation which
imposes a large-$N_c$ limit, we have also access to the Gaussian truncation
(GT) which, although sharing the dynamical content of the BK-equation, offers
a more refined prescription to map this information onto correlators. This
includes correlators that are not accessible in the BK-approximation itself
and leads to a closer match with JIMWLK evolution, see~\cite{Kovchegov:2008mk}
for a detailed discussion. The exposition below first recapitulates the main
steps in deriving $q\Bar q$ and $q\Bar q g$-correlators already given there
and in~\cite{Weigert:2005us} and extends these results to the four Wilson line
correlators needed to address inclusive vector-meson production.

The starting point is a parametrization of JIMWLK-averages in terms of the
functional prescription
\begin{align}
  \label{eq:func-Gaussian-average}
  \langle \ldots \rangle_Y = \exp\biggl\{ 
    -\frac12 \int\limits^Y dY' \int\! d^2x\, d^2y\ 
    G_{Y',\bm{x y}}\ \frac{\delta}{\delta A^{a +}_{\bm x,Y'}}
    \frac{\delta}{\delta A^{a +}_{\bm y,Y'}}
  \biggr\} \ldots
\ ,
\end{align}
where the dots ``$\ldots$'' stands for some generic $U$-field correlator such
as $\tr(U_{\bm x} U^\dagger_{\bm y})$, $U_{\bm z}^{a b} \tr(t^a U_{\bm x}t^b
U^\dagger_{\bm y})$ or the correlators in Eq.~\eqref{eq:qqb2-channel-matrix}.
In the light-cone gauge $A^-=0$, the Wilson lines depend only on the $A^+$
component of the target gluon field, and this gauge choice allows to write
Eq.~\eqref{eq:func-Gaussian-average} in terms of $\delta/\delta A^+$
explicitely.  We have done so for simplicity, but it is not necessary.  In
App.~\ref{sec:link-with-non-local-gauss}, we show that this parametrization of
the target averages $\langle \ldots \rangle_Y$ is equivalent to parametrizing
$W_Y[\rho]$ as a non-local Gaussian, as proposed
in~\cite{Fujii:2006ab,Marquet:2007vb}.

This recasts JIMWLK-evolution in terms of a single $Y$-dependent two point
function $G$, which represents two gluon exchanges between projectile and
target.

Its $Y$-dependence can be determined from the Balitsky evolution equation for
a $q\Bar q$- or more generally a generic ${\cal R}\Bar{\cal R}$-correlator
(with ${\cal R}$ denoting a generic representation),
  \begin{align}
    \label{eq:prefactUR}
    \frac{d}{dY} 
 \bigl\langle 
\overset{{\scriptscriptstyle\cal R}}\tr(
\overset{{\scriptscriptstyle\cal R}} U_{\bm{x}}
\overset{{\scriptscriptstyle\cal R}}
  U^\dagger_{\bm{y}}) 
\bigr\rangle_Y    
=\frac{\alpha_s}{\pi^2}\int\! d^2z\ {\cal K}_{\bm{x z y}} \
  \biggl(
  \bigl\langle
\big[\Tilde U_{\bm{z}}\big]^{a b}\
   \overset{{\scriptscriptstyle\cal R}}\tr(
\overset{{\scriptscriptstyle\cal R}} t^a 
\overset{{\scriptscriptstyle\cal R}} U_{\bm{x}}
\overset{{\scriptscriptstyle\cal R}} t^b 
\overset{{\scriptscriptstyle\cal R}}
  U^\dagger_{\bm{y}})
\bigr\rangle_Y 
  -C_{\cal R}  
\bigl\langle
\overset{{\scriptscriptstyle\cal R}}\tr(
\overset{{\scriptscriptstyle\cal R}} U_{\bm{x}}
\overset{{\scriptscriptstyle\cal R}}
  U^\dagger_{\bm{y}})  \bigr\rangle_Y    
\biggr)
\ ,
\end{align}
once $\bigl\langle \overset{{\scriptscriptstyle\cal R}}\tr(
\overset{{\scriptscriptstyle\cal R}} U_{\bm{x}}
\overset{{\scriptscriptstyle\cal R}} U^\dagger_{\bm{y}}) \bigr\rangle_Y$ and
$\bigl\langle \big[\Tilde U_{\bm{z}}\big]^{a b}\
\overset{{\scriptscriptstyle\cal R}}\tr( \overset{{\scriptscriptstyle\cal R}}
t^a \overset{{\scriptscriptstyle\cal R}} U_{\bm{x}}
\overset{{\scriptscriptstyle\cal R}} t^b \overset{{\scriptscriptstyle\cal R}}
U^\dagger_{\bm{y}}) \bigr\rangle_Y$ are known in terms of $G$,
see~\cite{Kovchegov:2008mk,Weigert:2005us} and the discussion
below.\footnote{If ${\cal R}$ is the fundamental
  representation,~\eqref{eq:trUUdagger-app} gives rise to the BK equation in
  the large-$N_c$ limit.}

The result is still a truncation of the Balitsky hierarchies (and thus
JIMWLK-evolution) in that the Balitsky equation for a general three point
function (for example the one already present on the
r.h.s. of~\eqref{eq:prefactUR}) would impose additional conflicting
constraints on $G$ -- this can be only overcome by introducing additional
degrees of freedom in form of higher $n$-point functions $G_{Y; \bm x_1\ldots\bm
  x_n}$ into the functional~\eqref{eq:func-Gaussian-average} to step beyond
the Gaussian truncation. 

Once such higher end point functions are crucial to the physics content of an
observable, such generalizations might become phenomenologically
indispensable. Still, the Gaussian truncation will yield an important
``baseline'' contribution to which the higher $n$-point functions $G_{Y; \bm
  x_1\ldots\bm x_n}$ provide corrections and must, therefore, be understood
first.

Staying within the Gaussian truncation,
the $U$-field correlators can generally be expressed in terms of 
\begin{align}
  \label{eq:calGdef}
  {\cal G}_{Y,\bm{x y}} := \int\limits^Y\! dY'\biggr(G_{Y',\bm{x y}}
    -\frac12\bigl(G_{Y',\bm{x x}}+G_{Y',\bm{y y}}\bigr)\biggr)
\end{align}
or its $Y$-derivative ${\cal G}'_{Y,\bm{x y}}$. 

The equation for ${\cal G}$ has already been derived in~\cite{Weigert:2005us}
(and in a somewhat different form earlier in~\cite{Kovner:2001vi}\footnote{To
  connect with the form given in~\cite{Kovner:2001vi}, Eq.~(5.3), one should
  reconstruct the evolution equation for the $q\Bar q$-dipole operator by
  multiplying~(\ref{eq:tilde-G-evo-short}) with $\exp(-C_f {\cal G}_{Y,\bm{x
      y}})$ and note that our ${\cal G}_{Y,\bm{x y}}$ corresponds to $v(\bm
  x,\bm y)$ in~\cite{Kovner:2001vi}.  }), starting from the $q\Bar q$-dipole
evolution equation~\eqref{eq:prefactUR} with ${\cal R}$ chosen to be the
fundamental representation. In~\cite{Kovchegov:2008mk} it was shown
that~\eqref{eq:func-Gaussian-average} treats {\em all} dipole equations
consistently: Applying~\eqref{eq:func-Gaussian-average} to the generic dipole
evolution equation~(\ref{eq:prefactUR}) yields one and the same equation for
${\cal G}$,
\begin{align}
  \label{eq:tilde-G-evo-short}
 \frac{d}{d Y} {\cal G}_{Y,\bm{x y}} & =  \frac{\alpha_s}{\pi^2} \int\!\!d^2z\  
 {\cal K}_{\bm{x z y}} \biggl(
 1-  e^{-\frac{ N_c}{2} \bigl(
 {\cal G}_{Y,{\bm{x z}}} +{\cal G}_{Y,{\bm{y z}}}
 - {\cal G}_{Y,{\bm{x y}}}\bigr)}
\biggr)
\ ,
\end{align}
{\em irrespective} of the representation ${\cal R}$.

The dynamical information driving the evolution in the Gaussian truncation is
in fact equivalent to that of the BK equation~\cite{Kovchegov:2008mk}. As a
consequence the kernel in Eq.~\eqref{eq:tilde-G-evo-short} is in fact the
BK-kernel
\begin{align}
  \label{eq:BK-kernel}
{\cal K}_{\bm{x z y}} & = 
    \frac{(\bm x-\bm y)^2}{
      (\bm x-\bm z)^2(\bm z-\bm y)^2
    }
\ .
\end{align}
Like the solutions to the BK equation, the solutions to
Eq.~(\ref{eq:tilde-G-evo-short}) generically approch the black disk limit with
${\cal G}_{Y,\bm x\bm y}\to\infty$ at $|\bm x-\bm y|\to\infty$.

It is possible to extend the result to NLO either by using the full results
of~\cite{Balitsky:2008zz, Balitsky:2008rc, Balitsky:2009xg, Balitsky:2009yp}
supplemented with a resummation of running coupling corrections, or, as
briefly outlined and used phenomenologically in~\cite{Weigert:2009zz,
  HW-DIS2009}, by combining the resummed running coupling results with
``energy conservation corrections''~\cite{Gotsman:2004xb} to summarize the
conformal NLO contributions in a numerically advantageous manner.

The Gaussian-truncation improves upon the BK-truncation in the way it maps
this information onto correlators: it allows to calculate consistent
expressions for any Wilson line correlator, even those subleading in a
$1/N_c$-expansion. The correlators obtained
from~\eqref{eq:func-Gaussian-average} automatically respect group theory
constraints that are inherited from field level relationships in various
coincidence limits, such as
\begin{subequations}
  \label{eq:3-point-generic-coincidence}
\begin{align}
  \label{eq:generic-x=y-limit}
\lim\limits_{y\to x}     \big[\Tilde U_{\bm{z}}\big]^{a b}\
   \overset{{\scriptscriptstyle\cal R}}\tr(
\overset{{\scriptscriptstyle\cal R}} t^a 
\overset{{\scriptscriptstyle\cal R}} U_{\bm{x}}
\overset{{\scriptscriptstyle\cal R}} t^b 
\overset{{\scriptscriptstyle\cal R}}
  U^\dagger_{\bm{y}})
& =  C_{\cal R} \frac{d_{\cal R}}{d_A}
\Tilde\tr\left(\Tilde U_{\bm z} \Tilde U_{\bm x}^\dagger \right)
\ ,
\\
  \label{eq:z=x,y-limit}
\lim\limits_{
      \bm z\to \bm y\ \text{or}\ \bm x}    
\big[\Tilde U_{\bm{z}}\big]^{a b}\
   \overset{{\scriptscriptstyle\cal R}}\tr(
\overset{{\scriptscriptstyle\cal R}} t^a 
\overset{{\scriptscriptstyle\cal R}} U_{\bm{x}}
\overset{{\scriptscriptstyle\cal R}} t^b 
\overset{{\scriptscriptstyle\cal R}}
  U^\dagger_{\bm{y}}) 
& = C_{\cal R}   \,
\overset{{\scriptscriptstyle\cal R}}\tr(
\overset{{\scriptscriptstyle\cal R}} U_{\bm{x}}
\overset{{\scriptscriptstyle\cal R}} U_{\bm{y}}^\dagger
) 
\ ,
\\
\lim\limits_{
      \bm z\to \bm y;\ \bm y\to \bm x}    
\big[\Tilde U_{\bm{z}}\big]^{a b}\
   \overset{{\scriptscriptstyle\cal R}}\tr(
\overset{{\scriptscriptstyle\cal R}} t^a 
\overset{{\scriptscriptstyle\cal R}} U_{\bm{x}}
\overset{{\scriptscriptstyle\cal R}} t^b 
\overset{{\scriptscriptstyle\cal R}}
  U^\dagger_{\bm{y}}) 
& = C_{\cal R} \, d_{\cal R}
\ ,
\end{align}
\end{subequations}
to use the correlators entering~\eqref{eq:prefactUR} as an example.
[$d_{\cal R}$ stands for the dimension of the representation
($d_f=N_c$ for the fundamental representation, $d_A=N_c^2-1$ for
adjoint, etc.) and $\Tilde\tr$ denotes the trace in the adjoint
representation.] 

Such coincidence limits are built into full JIMWLK-evolution, but are true in
the BK-truncation only in the leading $N_c$ sense. This is the main reason for
the closer match between full JIMWLK-evolution and GT compared to BK observed
in\cite{Kovchegov:2008mk}. For instance, with ${\cal R}$ chosen to be the
fundamental representation, the coincidence limit \eqref{eq:generic-x=y-limit}
used together with the Fierz identity
\begin{equation}
  \label{eq:Fierz}
  \big[\Tilde U_{\bm{z}}\big]^{a b}
  2 \tr(t^a U_{\bm{x}}t^b U^\dagger_{\bm{y}}) =
  \tr( U_{\bm{x}}U^\dagger_{\bm{z}}) \
           \tr( U_{\bm{z}}U^\dagger_{\bm{y}})
           -\frac{1}{N_c}\tr( U_{\bm{x}}U^\dagger_{\bm{y}})
\ ,
\end{equation}
yields the relation
\begin{equation}
  \label{eq:adj-fun}
  \Tilde\tr\left(\Tilde U_{\bm z} \Tilde U_{\bm x}^\dagger \right)=
  \left|\tr( U_{\bm{x}}U^\dagger_{\bm{z}})\right|^2-1
\ .
\end{equation}
This relates fundamental and adjoint dipole operators at the field level,
prior to target averaging.  Performing this averaging in the BK approximation,
one finds that Eq.~\eqref{eq:adj-fun} is only true in the leading $1/N_c$
sense.  By contrast, all group theory contstraints remain \emph{exact} in the
Gaussian truncation, and in this respect evolution in the GT approximation is
closer to full JIMWLK evolution. In the black-disk regime, one has
$\langle\Tilde\tr(\Tilde U_{\bm y} \Tilde U_{\bm x}^\dagger)\rangle_Y=0$, and
therefore Eq.~\eqref{eq:adj-fun} imposes that $\langle\left|\tr(
  U_{\bm{x}}U^\dagger_{\bm{y}})\right|^2\rangle_Y=1$ in the fully saturated
regime, meaning $\langle|\Hat S_{\bm x\bm y}^{q\Bar q}|^2\rangle_Y=1/N_c^2$.
The result remains finite as required by group theory -- a subtle feature that
is lost in the BK-approximation.

The Gaussian truncation is not the only approximation proposed to go beyond
BK.  In the large-$N_c$ limit, only dipole degrees of freedom are left in the
JIMWLK evolution, meaning that Wilson lines can enter only through the
operator $\Hat S_{\bm x\bm y}^{q\Bar q}$.  In this context, a family a
solutions was found in Ref.~\cite{Janik:2004ts,Janik:2004ve}. It is
parametrized by a real parameter $c$, with values between zero and one. The
treatment goes beyond the BK approximation as it does not assume correlator
factorization and solves the whole hierarchy of equations for $n$-$\Hat S_{\bm
  x\bm y}^{q\Bar q}$ operators emerging from JIMWLK evolution in this
approximation.  The solutions of Ref.~\cite{Janik:2004ts,Janik:2004ve} take
the form $\langle\Hat S_{\bm x\bm y}^{q\Bar q}\rangle^c_Y=1-cT(\bm x,\bm y;Y)$
and $\langle\Hat S_{\bm y'\bm x'}^{q\Bar q}\Hat S_{\bm x\bm y}^{q\Bar
  q}\rangle^c_Y -\langle \Hat S_{\bm y'\bm x'}^{q\Bar q}\rangle_Y^c\langle\Hat
S_{\bm x\bm y}^{q\Bar q}\rangle^c_Y =c(1-c)T(\bm x,\bm y;Y)T(\bm x',\bm
y';Y)$, where $1-T(\bm x,\bm y;Y)$ is obtained from the (factorized) BK
equation (see~\cite{Janik:2004ts,Janik:2004ve} for details). As a consequence
it leads to a ``gray disk limit'' in the sense that $\langle\Hat S_{\bm x\bm
  y}^{q\Bar q}\rangle^c_Y\to 1-c$ at $|\bm x-\bm y|\to\infty$ instead of zero,
the ``black disk limit.''

By contrast, the GT approximation does not assume the large-$N_c$ limit, but
solves only the first equation of the hierarchy.  The correlations obtained in
the GT approximation are more complex and always satisfy group theory
constraints \emph{exactly} (c.f.~\eqref{eq:3-point-generic-coincidence})
instead of up to terms subleading in $1/N_c$.  In addition it always leads to
$\langle\Hat S_{\bm x\bm y}^{q\Bar q}\rangle_Y\to 0$ at $|\bm x-\bm
y|\to\infty$ i.e. the black disk limit. For the initial conditions considered
in~\cite{Rummukainen:2003ns, Kovchegov:2008mk} agreement with
JIMWLK-simulations is excellent.

\subsection{Efficient construction of correlators}
\label{sec:efficient-construction-of-correlators}

To address the four Wilson line correlator encountered in inclusive
vector-meson production, one has to step beyond the correlators already
obtained in~\cite{Kovchegov:2008mk} -- to prepare for that it helps to review
how this was achieved with a number of simple examples.

While direct calculation of specific correlators
using the averaging procedure~(\ref{eq:func-Gaussian-average}) is in many
cases straightforward, the calculation can be often simplified significantly
by using differential equations.  This relies on the observation that
\begin{align}
  \label{eq:gen-dY-eq-app}
  \frac{d}{dY}\langle \ldots \rangle_Y =\ & -\frac12 \langle \int\!
  d^2u\, d^2v\ G_{Y,\bm{u v}} \frac{\delta}{\delta A^{a +}_{\bm u,Y}}
  \frac{\delta}{\delta A^{a +}_{\bm v,Y}} \ldots \rangle_Y 
\ ,
\end{align}
which can be used to advantage if the right-hand side turns out to be in some
form proportional to $\langle \ldots \rangle(Y)$ itself.  Note that the
derivatives $\delta/\delta A^{a +}_{\bm x,Y}$ in~\eqref{eq:gen-dY-eq-app}
--contrary to those in~\eqref{eq:func-Gaussian-average}-- only touch the
largest value of $x^-$ when acting on the $U$'s, which are $x^-$-ordered
exponentials of $t^aA^{a +}$.  Our convention is that larger values of $x^-$
appear further to the left in a $U$, and therefore further to the right in a
$U^\dagger$.

\paragraph{Two point projectile ${\cal R}$-$\bar {\cal R}$ correlators:}
Using the notation~(\ref{eq:calGdef}) and a prime to denote a $Y$
derivative, straightforward algebra leads to
\begin{align}
  \label{eq:trUUdagger-app}
   \frac{d}{dY}\langle \overset{{\scriptscriptstyle\cal R}}\tr(
\overset{{\scriptscriptstyle\cal R}} U_{\bm{x}}
\overset{{\scriptscriptstyle\cal R}}
  U^\dagger_{\bm{y}})
\rangle_Y
= -{\cal G}'_{Y,\bm{x y}}\ 
\langle \overset{{\scriptscriptstyle\cal R}}\tr(
\overset{{\scriptscriptstyle\cal R}} t^a 
\overset{{\scriptscriptstyle\cal R}} U_{\bm{x}}
\overset{{\scriptscriptstyle\cal R}}
  U^\dagger_{\bm{y}}\overset{{\scriptscriptstyle\cal R}} t^a )
\rangle_Y
= -C_{\cal R} {\cal G}'_{Y,\bm{x y}}\ 
\langle \overset{{\scriptscriptstyle\cal R}}\tr(
\overset{{\scriptscriptstyle\cal R}} U_{\bm{x}}
\overset{{\scriptscriptstyle\cal R}}
  U^\dagger_{\bm{y}})
\rangle_Y
\end{align}
which is readily solved to obtain
\begin{align}
  \label{eq:UUdaggersol-app}
  \langle \overset{{\scriptscriptstyle\cal R}}\tr(
\overset{{\scriptscriptstyle\cal R}} U_{\bm{x}}
\overset{{\scriptscriptstyle\cal R}}
  U^\dagger_{\bm{y}})
\rangle_Y = d_{\cal R} e^{-C_{\cal R}{\cal G}_{Y,\bm{x y}}}
\ . 
\end{align}
The freedom in the initial condition was used to accommodate the normalization
factor $d_{\cal R}$.

\paragraph{Three point projectile adjoint-${\cal R}$-$\bar {\cal R}$ correlators:}
These involve several distinct color structures.
\begin{align}
  \label{eq:UtrtUtUdagger-eq-app}
   \frac{d}{dY} & \langle 
\big[\Tilde U_{\bm{z}}\big]^{a b} 
\overset{{\scriptscriptstyle\cal R}}\tr(
\overset{{\scriptscriptstyle\cal R}} t^a 
\overset{{\scriptscriptstyle\cal R}} U_{\bm{x}}
\overset{{\scriptscriptstyle\cal R}} t^b 
\overset{{\scriptscriptstyle\cal R}}
  U^\dagger_{\bm{y}})
\rangle_Y
=
-{\cal G}'_{Y,\bm{x z}} \langle 
\big[\Tilde t^i \Tilde U_{\bm{z}}\big]^{a b}\
\overset{{\scriptscriptstyle\cal R}}\tr(
\overset{{\scriptscriptstyle\cal R}} t^a 
\overset{{\scriptscriptstyle\cal R}} t^i 
\overset{{\scriptscriptstyle\cal R}} U_{\bm{x}}
\overset{{\scriptscriptstyle\cal R}} t^b 
\overset{{\scriptscriptstyle\cal R}}
  U^\dagger_{\bm{y}})
\rangle_Y
\notag \\ & \
+
{\cal G}'_{Y,\bm{z y}} \langle 
\big[\Tilde t^i \Tilde U_{\bm{z}}\big]^{a b}\
\overset{{\scriptscriptstyle\cal R}}\tr(
\overset{{\scriptscriptstyle\cal R}} t^a 
\overset{{\scriptscriptstyle\cal R}} U_{\bm{x}}
\overset{{\scriptscriptstyle\cal R}} t^b 
\overset{{\scriptscriptstyle\cal R}}
  U^\dagger_{\bm{y}} \overset{{\scriptscriptstyle\cal R}} t^i 
)
\rangle_Y
+
{\cal G}'_{Y,\bm{x y}}
\langle 
\big[\Tilde U_{\bm{z}}\big]^{a b}\
\overset{{\scriptscriptstyle\cal R}}\tr(
\overset{{\scriptscriptstyle\cal R}} t^a 
\overset{{\scriptscriptstyle\cal R}} t^i 
\overset{{\scriptscriptstyle\cal R}} U_{\bm{x}}
\overset{{\scriptscriptstyle\cal R}} t^b 
\overset{{\scriptscriptstyle\cal R}}
  U^\dagger_{\bm{y}}
\overset{{\scriptscriptstyle\cal R}} t^i 
)
\rangle_Y
\notag \\
\notag \\
= \ &
-\left[\frac{N_c}2\left(
  {\cal G}'_{Y,\bm{x z}} + {\cal G}'_{Y,\bm{z y}} 
\right)
+ {\cal G}'_{Y,\bm{x y}} \left(C_{\cal R}-\frac{N_c}2\right)
\right]
\langle 
\big[\Tilde U_{\bm{z}}\big]^{a b}\
\overset{{\scriptscriptstyle\cal R}}\tr(
\overset{{\scriptscriptstyle\cal R}} t^a 
\overset{{\scriptscriptstyle\cal R}} U_{\bm{x}}
\overset{{\scriptscriptstyle\cal R}} t^b 
\overset{{\scriptscriptstyle\cal R}}
  U^\dagger_{\bm{y}})
\rangle_Y
\end{align}
where we have used
\begin{align}
  \label{eq:t-resuffle-1-app}
\overset{{\scriptscriptstyle\cal R}} t^i 
\overset{{\scriptscriptstyle\cal R}} t^a 
\overset{{\scriptscriptstyle\cal R}} t^i & = 
\overset{{\scriptscriptstyle\cal R}} t^a 
\overset{{\scriptscriptstyle\cal R}} t^i 
\overset{{\scriptscriptstyle\cal R}} t^i 
+
[\overset{{\scriptscriptstyle\cal R}} t^i
, 
\overset{{\scriptscriptstyle\cal R}} t^a 
]
\overset{{\scriptscriptstyle\cal R}} t^i \notag \\
& = 
\overset{{\scriptscriptstyle\cal R}} t^a 
C_{\cal R} 
+ i f_{i a k}
\overset{{\scriptscriptstyle\cal R}} t^k 
\overset{{\scriptscriptstyle\cal R}} t^i
= \overset{{\scriptscriptstyle\cal R}} t^a 
C_{\cal R} 
+ i f_{i a k}
\frac12 i f_{k i j} \overset{{\scriptscriptstyle\cal R}} t^j
=  
\overset{{\scriptscriptstyle\cal R}} t^a \left(C_{\cal R}-\frac{N_c}2\right).
\end{align}
The nontrivial point here is, that this holds for any representation ${\cal
  R}$.  Integrating~(\ref{eq:UtrtUtUdagger-eq-app}) one finds
\begin{align}
  \label{eq:UtrtUtUdagger-app}
\langle 
\big[\Tilde U_{\bm{z}}\big]^{a b} &
\overset{{\scriptscriptstyle\cal R}}\tr(
\overset{{\scriptscriptstyle\cal R}} t^a 
\overset{{\scriptscriptstyle\cal R}} U_{\bm{x}}
\overset{{\scriptscriptstyle\cal R}} t^b 
\overset{{\scriptscriptstyle\cal R}}
  U^\dagger_{\bm{y}})
\rangle_Y = C_{\cal R} d_{\cal R} e^{
  -\frac{N_c}2\left(
  {\cal G}_{Y,\bm{x z}} + {\cal G}_{Y,\bm{z y}}- {\cal G}_{Y,\bm{x y}}  
\right)
-C_{\cal R} {\cal G}_{Y,\bm{x y}} 
}  
\ ,
\end{align}
again with the free initial condition used to set the normalization properly.

\section{Four Wilson lines in the Gaussian truncation: correlators for
  inclusive vector-meson production}
\label{sec:quadr-evol-gauss}

To address inclusive vector-meson production in the Gaussian truncation, one
needs to step beyond two- and three-point functions and to derive an
expression for the four point correlator $\langle \Hat S_{\bm y'\bm x'}^{q\Bar
  q}\Hat S_{\bm x\bm y}^{q\Bar q}\rangle_Y$ in Eq.~\eqref{eq:qqb2-corr-gen}.

As it turns out, this does not take the form of a simple exponentiation as the
examples encountered earlier. The reason is the richer color structure
associated with the four Wilson lines: with two quarks and two anti-quarks
available, one can form not only one, but two non-equivalent singlets, which
will mix under evolution in $Y$. Where~\eqref{eq:qqb2-corr-gen} forms a singlet
directly from $q\Bar q$ pairs, one may equally well have each of them in an
octet state which then together form a singlet -- transition elements between
the two singlets in general are also non-vanishing. One is led to consider a
$2\times 2$-matrix of correlators, of which only the $(1,1)$-component enters
the meson-production cross-sections of
Sec.~\ref{sec:Wilson-lines-beyond-total-cross} directly:
\begin{align}
  \label{eq:qqb2-channel-matrix}
{\cal A}(Y)  := &\, 
  \begin{pmatrix}
\frac{
\langle
\tr(U_{\bm y'} U_{\bm x'}^\dagger) 
\tr(U_{\bm x} U_{\bm y}^\dagger)
\rangle_Y
}{N_c^2}
\vphantom{
\diagram[height=1.3cm]{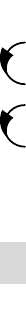}
\diagram[height=1.3cm]{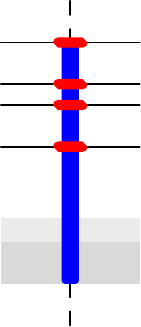}
\diagram[height=1.3cm]{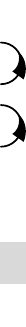}}
&  
\frac{
\langle
\tr(U_{\bm y'} t^a U_{\bm x'}^\dagger) 
\tr(U_{\bm x} t^a U_{\bm y}^\dagger)
\rangle_Y
}{N_c \sqrt{d_A/4}}
\vphantom{
\diagram[height=1.3cm]{qqbqqbsst-L}
\diagram[height=1.3cm]{qqbqqbintreg}
\diagram[height=1.3cm]{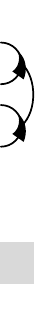}}
\\
\frac{
\langle 
\tr(U_{\bm y'} U_{\bm x'}^\dagger t^b ) 
\tr(U_{\bm x} U_{\bm y}^\dagger t^b )
\rangle_Y
}{N_c \sqrt{d_A/4}}
\vphantom{
\diagram[height=1.3cm]{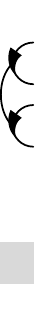}
\diagram[height=1.3cm]{qqbqqbintreg}
\diagram[height=1.3cm]{qqbqqbsst-R}
}
& 
\frac{
\langle \tr(U_{\bm y'}  t^a U_{\bm x'}^\dagger t^b ) 
\tr(U_{\bm x} t^a U_{\bm y}^\dagger t^b )
\rangle_Y
}{d_A/4}
\vphantom{
\diagram[height=1.3cm]{qqbqqboost-L}
\diagram[height=1.3cm]{qqbqqbintreg}
\diagram[height=1.3cm]{qqbqqboost-R}
}
\end{pmatrix}
\notag \\  = &\,
  \begin{pmatrix}
\frac1{\diagram[width=.11cm]{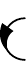}\diagram[width=.11cm]{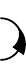}^2}
\diagram[height=1.3cm]{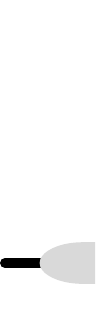}
\diagram[height=1.3cm]{qqbqqbsst-L}
\diagram[height=1.3cm]{qqbqqbintreg-R}
\diagram[height=1.3cm]{qqbqqbsst-R}
\diagram[height=1.3cm]{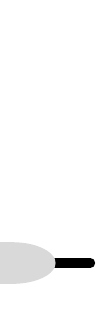}
&  
\frac1{\diagram[width=.11cm]{qqbs-L}\diagram[width=.11cm]{qqbs-R}} 
\frac1{\diagram[width=.11cm]{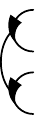}\diagram[width=.11cm]{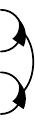}^{\frac12}}
\diagram[height=1.3cm]{t2wf-L}
\diagram[height=1.3cm]{qqbqqbsst-L}
\diagram[height=1.3cm]{qqbqqbintreg-R}
\diagram[height=1.3cm]{qqbqqboost-R}
\diagram[height=1.3cm]{t2wf-R}
\\
 \frac1{\diagram[width=.11cm]{qqbs-L}\diagram[width=.11cm]{qqbs-R}} 
\frac1{\diagram[width=.11cm]{qqbqqboos-L}\diagram[width=.11cm]{qqbqqboos-R}^{\frac12}}
\diagram[height=1.3cm]{t2wf-L}
\diagram[height=1.3cm]{qqbqqboost-L}
\diagram[height=1.3cm]{qqbqqbintreg-R}
\diagram[height=1.3cm]{qqbqqbsst-R}
\diagram[height=1.3cm]{t2wf-R}
& 
\frac1{\diagram[width=.11cm]{qqbqqboos-L}\diagram[width=.11cm]{qqbqqboos-R}}
\diagram[height=1.3cm]{t2wf-L}
\diagram[height=1.3cm]{qqbqqboost-L}
\diagram[height=1.3cm]{qqbqqbintreg-R}
\diagram[height=1.3cm]{qqbqqboost-R}
\diagram[height=1.3cm]{t2wf-R}
\end{pmatrix}(Y)
\end{align}
where $d_A = N_c^2-1$ is the dimension of the adjoint representation. The
diagrammatic representation in the second line employs standard birdtrack
notation to clarify the color structures (see~\cite{Cvitanovic-GroupTheory}
for a textbook introduction). The decomposition of a $q\Bar q$
state into singlet and octet is written as
\begin{align}
  \label{eq:octett+singlett=id-flat-direct}
  \diagram[height=.65cm]{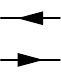} = & 
\frac1{\diagram[width=.15cm]{qqbs-L}\diagram[width=.15cm]{qqbs-R}} 
 \diagram[height=.65cm]{qqbs-R}\diagram[height=.65cm]{qqbs-L}
+  2\ \diagram[height=.6cm]{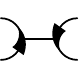} 
\ ,
\end{align}
and used to form a color basis to span the space of two non-equivalent
singlets according to
\begin{equation}
  \label{eq:qqbqqb-singlets-1}
  \frac1{\diagram[width=.15cm]{qqbs-L}\diagram[width=.15cm]{qqbs-R}}
  \diagram[height=.8cm]{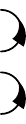}
  \hspace{1cm}\text{and} \hspace{1cm}
  \frac1{\diagram[width=.15cm]{qqbqqboos-L}
    \diagram[width=.15cm]{qqbqqboos-R}^{\frac{1}{2}}} 
  \diagram[height=.8cm]{qqbqqboos-R}
\end{equation}
with normalization factors given by
\begin{equation*}
    \diagram[width=.15cm]{qqbs-L}
  \diagram[width=.15cm]{qqbs-R} = N_c, 
  \hspace{1cm}\text{and} \hspace{1cm}
  \diagram[height=.5cm]{qqbqqboos-L}
  \diagram[height=.5cm]{qqbqqboos-R} 
  =
  \diagram[height=.5cm]{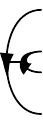}
  \diagram[height=.5cm]{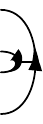} 
  = 
  \frac14(N_c^2-1)= \frac{d_A}4
\ ,
\end{equation*}
with the $\frac14$ related to the the normalization of generators $\tr(t^a
t^b) = \frac12 \delta^{a b}$.

This notation facilitates a more immediate structural identification of the
meaning of the analytical expressions given in the first line of
Eq.~\eqref{eq:qqb2-channel-matrix}.

The reason one can not simply focus on ${\cal A}_{1 1}(Y)$ to the exclusion of
the remaining three components is that, in general, they mix under
JIMWLK-evolution even if the off diagonal elements
in~\eqref{eq:qqb2-channel-matrix} vanish at some $Y_0$. To see this, one notes
that already the linear contributions to JIMWLK-evolution (the terms in
Eq.~\eqref{eq:qqbqqb-amp-JIMWLK} with non-interacting gluons), such as
\begin{align}
  \label{eq:calA-evo-mix-rr}
  \begin{pmatrix}
\frac1{\diagram[width=.11cm]{qqbs-L}\diagram[width=.11cm]{qqbs-R}^2}
\diagram[height=1.1cm]{t2wf-L}\diagram[height=1.1cm]{qqbqqbsst-L}
\diagram[height=1.1cm]{qqbqqbintreg-rr-R}
\diagram[height=1.1cm]{qqbqqbsst-R}
\diagram[height=1.1cm]{t2wf-R}
&  
\frac1{\diagram[width=.11cm]{qqbs-L}\diagram[width=.11cm]{qqbs-R}} 
\frac1{\diagram[width=.11cm]{qqbqqboos-L}\diagram[width=.11cm]{qqbqqboos-R}^{\frac12}}
\diagram[height=1.1cm]{t2wf-L}\diagram[height=1.1cm]{qqbqqbsst-L}
\diagram[height=1.1cm]{qqbqqbintreg-rr-R}
\diagram[height=1.1cm]{qqbqqboost-R}
\diagram[height=1.1cm]{t2wf-R}
\\
 \frac1{\diagram[width=.11cm]{qqbs-L}\diagram[width=.11cm]{qqbs-R}} 
\frac1{\diagram[width=.11cm]{qqbqqboos-L}\diagram[width=.11cm]{qqbqqboos-R}^{\frac12}}
\diagram[height=1.1cm]{t2wf-L}\diagram[height=1.1cm]{qqbqqboost-L}
\diagram[height=1.1cm]{qqbqqbintreg-rr-R}
\diagram[height=1.1cm]{qqbqqbsst-R}
\diagram[height=1.1cm]{t2wf-R}
& 
\frac1{\diagram[width=.11cm]{qqbqqboos-L}\diagram[width=.11cm]{qqbqqboos-R}}
\diagram[height=1.1cm]{t2wf-L}\diagram[height=1.1cm]{qqbqqboost-L}
\diagram[height=1.1cm]{qqbqqbintreg-rr-R}
\diagram[height=1.1cm]{qqbqqboost-R}
\diagram[height=1.1cm]{t2wf-R}
\end{pmatrix}(Y)
= {\sf M} \circ   \begin{pmatrix}
\frac1{\diagram[width=.11cm]{qqbs-L}\diagram[width=.11cm]{qqbs-R}^2}
\diagram[height=1.1cm]{t2wf-L}
\diagram[height=1.1cm]{qqbqqbsst-L}
\diagram[height=1.1cm]{qqbqqbintreg-R}
\diagram[height=1.1cm]{qqbqqbsst-R}
\diagram[height=1.1cm]{t2wf-R}
&  
\frac1{\diagram[width=.11cm]{qqbs-L}\diagram[width=.11cm]{qqbs-R}} 
\frac1{\diagram[width=.11cm]{qqbqqboos-L}\diagram[width=.11cm]{qqbqqboos-R}^{\frac12}}
\diagram[height=1.1cm]{t2wf-L}
\diagram[height=1.1cm]{qqbqqbsst-L}
\diagram[height=1.1cm]{qqbqqbintreg-R}
\diagram[height=1.1cm]{qqbqqboost-R}
\diagram[height=1.1cm]{t2wf-R}
\\
 \frac1{\diagram[width=.11cm]{qqbs-L}\diagram[width=.11cm]{qqbs-R}} 
\frac1{\diagram[width=.11cm]{qqbqqboos-L}\diagram[width=.11cm]{qqbqqboos-R}^{\frac12}}
\diagram[height=1.1cm]{t2wf-L}
\diagram[height=1.1cm]{qqbqqboost-L}
\diagram[height=1.1cm]{qqbqqbintreg-R}
\diagram[height=1.1cm]{qqbqqbsst-R}
\diagram[height=1.1cm]{t2wf-R}
& 
\frac1{\diagram[width=.11cm]{qqbqqboos-L}\diagram[width=.11cm]{qqbqqboos-R}}
\diagram[height=1.1cm]{t2wf-L}
\diagram[height=1.1cm]{qqbqqboost-L}
\diagram[height=1.1cm]{qqbqqbintreg-R}
\diagram[height=1.1cm]{qqbqqboost-R}
\diagram[height=1.1cm]{t2wf-R}
\end{pmatrix}(Y)
\end{align}
couple the four components in ${\cal A}$, since ${\sf M}$ is a $2\times
2$-matrix with non-vanishing off diagonal elements. Its entries depend on all
four coordinates $\bm x$, $\bm y$, $\bm x'$, and $\bm y'$ present in ${\cal
  A}$ via the BK-kernel.\footnote{The structure of ${\sf M}$ can in fact be
  reconstructed from~\eqref{eq:qqb2-channel-gauss-diff}.} These linear terms
are the $q\Bar q$-analogues to the second term on the right hand side
of~\eqref{eq:prefactUR}.  The non-linear term (the analogue to the first term
on the right hand side of~\eqref{eq:prefactUR}) involves the entries of a
matrix of entirely new correlators in a space spanned by all singlets in a
$(q\Bar q)^2 g$ tensor product: there are altogether $6$ independent such
singlets, and therefore $6\times 6$ $(q\Bar q)^2 g$-correlators contributing
to the nonlinear term of full JIMWLK evolution
of~\eqref{eq:qqb2-channel-matrix}.

Even if one restricts oneself to the Gaussian truncation --and therefore
ignores the full complications of the nonlinear structure of the
JIMWLK-equation-- one can not escape the channel mixing already seen on the
linear level in Eq.~\eqref{eq:calA-evo-mix-rr}.

The correlator matrix ${\cal A}$ is hermitian even in the most general case of
four independent coordinates $\bm x$, $\bm y$, $\bm x'$ $\bm y'$, and thus can
in principle be diagonalized -- the diagonalizing transformation, however,
will in general be $Y$-dependent. This means that the procedure outlined in
Sec.~\ref{sec:efficient-construction-of-correlators}
to derive an expression for the four Wilson line correlator in the Gaussian
truncation leads to a solution in terms of a $Y$-ordered
exponential. Applying~\eqref{eq:gen-dY-eq-app}
to~\eqref{eq:qqb2-channel-matrix} (which chooses~\eqref{eq:qqbqqb-singlets-1}
as a basis) leads to a matrix equation of the form
\begin{align}
  \label{eq:qqb2-channel-gauss-diff}
  \frac{d}{d Y} {\cal A}(Y)
=   
-{\cal M}(Y)
{\cal A}(Y)
\end{align}
where matrix entries of ${\cal M}$ can be obtained by expanding ${\cal A}(Y)$
to first order in ${\cal G}'$. One finds
\begin{align}
  \label{eq:calMdef}
  {\cal M}(Y):=
\begin{pmatrix}
a_Y & c_Y \\
c_Y & b_Y
\end{pmatrix}
\end{align}
with (suppressing the $Y$-dependence on the ${\cal G}$ for compactness)
\begin{subequations}
  \label{eq:abc-values}
\begin{align}
  a_Y =\, & C_f 
  \Bigl({\cal G}'_{\bm x,\bm y}+{\cal G}'_{\bm x',\bm y'}\Bigr) \\
  b_Y =\, & \Bigl[ \Bigl(C_f-\frac{C_A}2\Bigr)
\Bigl({\cal G}'_{\bm x,\bm y} +{\cal G}'_{\bm x',\bm y'}\Bigr)
+
 \frac{C_d+C_A}4\Bigl( 
     {\cal G}'_{\bm x',\bm x} +
    {\cal G}'_{\bm y',\bm y} 
  \Bigr) 
-
\frac{C_d-C_A}4\Bigl( 
    {\cal G}'_{\bm x',\bm y} +
    {\cal G}'_{\bm y',\bm x}
    \Bigr) 
\Bigr]  \\
  c_Y =\, & \frac{\sqrt{d_A/4}}{C_A}
\Bigl(
  {\cal G}'_{\bm x',\bm x} 
+
  {\cal G}'_{\bm y',\bm y} 
-
  {\cal G}'_{\bm x',\bm y} 
-
  {\cal G}'_{\bm y',\bm x}
\Bigr)
\ .
\end{align}
\end{subequations} [$C_d= (N_c^2-4)/N_c$ is defined via the totally symmetric
tensors through $d^{a b c} d^{a b c'}= C_d \delta^{c c'}$ -- it vanishes at
$N_c=2$, where $d^{a b c}\to 0$ itself.]

The equation for ${\cal A}(Y)$ can be integrated to yield (with $P_Y$ denoting
path ordering in rapidity)
\begin{align}
  \label{eq:A-Y-Y-dep}
  {\cal A}(Y) = P_Y \exp\biggl[-\int\limits_{Y_0}^Y dY' 
\ {\cal M}(Y')
\biggr]
{\cal A}(Y_0)
\ ,
\end{align}
which relates ${\cal A}(Y_0)$ to ${\cal A}(Y)$ solely in terms of ${\cal G}'$,
the degrees of freedom in the Gaussian truncation. 

As it stands, the initial condition ${\cal A}(Y_0)$ can \emph{not} be
determined from that of a dipole correlator -- it contains new information
that can not be extracted from dipoles alone.  The freedom to choose ${\cal
  A}(Y_0)$ and its coordinate dependence is not unlimited, however, it is
rather strongly constrained by coincidence limits.

Direct inspection of Eq.~\eqref{eq:qqb2-channel-matrix} reveals that the
off-diagonal elements vanish when either $\bm y'\to \bm x'$ or $\bm y\to \bm
x$ (simply since the generators are traceless) in a $Y$-independent
manner. In these limits, the diagonal elements reduce to the two and three
point correlators encountered earlier, whose initial conditions contain no
freedom whatsoever, they are fully determined in terms of ${\cal G}(Y_0)$
alone. This then also determines ${\cal A}(Y_0)$ in this limit without further
freedom. The diagonal entries of ${\cal A}(Y_0)$ follow from
Eqns.~\eqref{eq:UUdaggersol-app} and~\eqref{eq:UtrtUtUdagger-app}, together
with the Fierz identity~\eqref{eq:Fierz}.

From the above, it is clear that if one swaps coordinates $\bm x'$ and $\bm y$
before forming singlets and octets, one obtains a matrix of correlators,
related to the above by using
\begin{equation}
  \label{eq:qqbqqb-singlets-2}
  \frac1{\diagram[width=.15cm]{qqbs-L}\diagram[width=.15cm]{qqbs-R}}
  \diagram[height=.8cm]{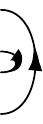}
  \hspace{1cm}\text{and} \hspace{1cm}
  \frac1{\diagram[width=.15cm]{qqbqqboos-L}
    \diagram[width=.15cm]{qqbqqboos-R}^{\frac{1}{2}}} 
  \diagram[height=.8cm]{qqbqqbthruoct-R}
\end{equation}
to replace~\eqref{eq:qqbqqb-singlets-1} via an orthogonal change of bases. In
this basis the limits $\bm x'\to\bm x$ and $\bm y'\to \bm y$ become evidently
diagonal at all $Y$, with the diagonal elements again fully determined by
known expressions in terms of ${\cal G}(Y)$. Again the diagonal entries of
${\cal A}(Y_0)$ in this set of limits can be directly read off from
Eqns.~\eqref{eq:UUdaggersol-app},~\eqref{eq:UtrtUtUdagger-app},
and~(\ref{eq:Fierz}).

The only set of pairwise coincidence limits not yet discussed is the situation
in which either both $U$-factors or both $U^\dagger$-factors are taken at the
same point.  $Y$-independent diagonalizability of ${\cal A}(Y)$ for this pair
of coincidence limits is exposed in a basis where one first decomposes both
quark- and anti-quark lines into symmetric and antisymmetric contributions that
then map into singlets. Details, including the limiting correlators on the
diagonal are given in App.~\ref{sec:decomp-four-wils-as-q2qb2}.

Evidently, this full set of coincidence limit constraints on ${\cal A}(Y_0)$
does not leave much quantitative freedom. The nontrivial freedom on the
initial condition is restricted to the configuration space regions away from
the coincidence limits, it resides in the difference between
path-ordered-exponentiation as in Eq.~\eqref{eq:A-Y-Y-dep} and ``rigid''
exponentiation as defined by\footnote{Note
  that the square roots in $\sf\Delta_Y$ are completely spurious -- the
  series representations of the expressions in Eq.~\eqref{eq:Exp_A_Y} strictly
  contain only integer powers of $\sf\Delta_Y$.}
\begin{align}
  \label{eq:Exp_A_Y}
 & {\cal A}^{\text{rigid}}(Y) := \exp\Bigl[ -
  \int^Y dY' {\cal M}(Y')
\Bigr]
=
\exp\Bigl[ -
  \begin{pmatrix}
    {\sf a}_Y & {\sf c}_Y \\
    {\sf c}_Y & {\sf b}_Y
  \end{pmatrix}
\Bigr]
=
  e^{-\frac12 ({\sf a}_Y+{\sf b}_Y)}
\notag \\ & \times
  \begin{pmatrix}
  \cosh\Bigl(\frac12\sqrt{\sf \Delta}_Y\Bigr)
  -\frac{{\sf a}_Y-{\sf b}_Y}{\sqrt{\sf \Delta}_Y} 
  \sinh\Bigl(\frac12\sqrt{\sf \Delta}_Y\Bigr) 
&
-2\frac{{\sf c}_Y }{\sqrt{\sf \Delta}_Y} 
  \sinh\Bigl(\frac12\sqrt{\sf \Delta}_Y\Bigr) 
\\
-2 \frac{{\sf c}_Y }{\sqrt{\sf \Delta}_Y} 
  \sinh\Bigl(\frac12\sqrt{\sf \Delta}_Y\Bigr) 
&
\cosh\Bigl(\frac12\sqrt{\sf \Delta}_Y\Bigr)
+\frac{{\sf a}_Y-{\sf b}_Y}{\sqrt{\sf \Delta}_Y} 
\sinh\Bigl(\frac12\sqrt{\sf \Delta}_Y\Bigr) 
  \end{pmatrix}
\end{align}
where
\begin{align}
  \label{eq:abc-int}
  {\sf a}_Y:=\int\limits^Y dY' a_{Y'} \hspace{1cm}
  {\sf b}_Y:=\int\limits^Y dY' b_{Y'} \hspace{1cm}
  {\sf c}_Y:=\int\limits^Y dY' c_{Y'} 
\end{align}
and 
\begin{align}
  \label{eq:Delta-def}
{\sf\Delta}_Y:= ({\sf a}_Y-{\sf b}_Y)^2+4 {\sf c}_Y^2  
\ .
\end{align}
${\cal A}^{\text{rigid}}(Y)$, contrary to ${\cal A}(Y)$, can be expressed
exclusively in terms of ${\cal G}(Y)$ instead of involving $Y$-ordered
expression in terms of ${\cal G}'(Y)$ and differs from ${\cal A}(Y)$ only away
from the coincidence limits.\footnote{The expression for $c_Y$ in each of the
  bases~\eqref{eq:qqbqqb-singlets-1}, \eqref{eq:qqbqqb-singlets-2},
  and~\eqref{eq:q2qb2-basis} vanishes by necessity in the associated pairwise
  limits. This carries over to its $Y$-integral ${\sf c_Y}$ and guarantees
  that ${\cal A}^{\text{rigid}}(Y)$ agrees with ${\cal A}(Y)$ in these
  limits.}

The large-$N_c$ limit is of no help in understanding the freedom left. It
simplifies the situation so drastically that all expressions for ${\cal A}(Y)$
in both the bases~\eqref{eq:qqbqqb-singlets-1}
and~\eqref{eq:qqbqqb-singlets-2} become completely diagonal and ordering plays
no role at all (c.f.  both~\eqref{eq:qqb2-channel-matrix}
and~\eqref{eq:abc-values}).\footnote{The third basis comes out to be
  non-diagonal, but by equivalence to the other two, has $Y$-independent
  eigenvectors.} The simplifications obtained, however, not only affect
configuration space away from the pairwise coincidence limits, also the result
for the limits looses subleading $1/N_c$ contributions correctly encoded into
the rigid exponentiation expression~\eqref{eq:Exp_A_Y} at finite $N_c$: The
diagonal entries in basis~\eqref{eq:qqbqqb-singlets-1} appear as the product
of two large-$N_c$ dipoles in keeping with Eq.~\eqref{eq:qqb2-corr-gen-Nc-exp}
\begin{align}
  \label{eq:calA-Nc-exp}
  {\cal A}(Y) = 
   \begin{pmatrix}
\frac{
\langle
\tr(U_{\bm y'} U_{\bm x'}^\dagger) 
\rangle_Y \langle
\tr(U_{\bm x} U_{\bm y}^\dagger)
\rangle_Y
}{N_c^2}
\vphantom{
\diagram[height=1.3cm]{qqbqqbsst-L}
\diagram[height=1.3cm]{qqbqqbintreg}
\diagram[height=1.3cm]{qqbqqbsst-R}}
&  
0
\\
0
& 
\frac{
\langle \tr(U_{\bm x} U_{\bm x'}^\dagger ) 
\rangle_Y \langle
\tr(U_{\bm y'} U_{\bm y}^\dagger )
\rangle_Y
}{N_c^2}
\vphantom{
\diagram[height=1.3cm]{qqbqqboost-L}
\diagram[height=1.3cm]{qqbqqbintreg}
\diagram[height=1.3cm]{qqbqqboost-R}
}
\end{pmatrix}
+{\cal O}(1/N_c)
\end{align}
with the two diagonal entries swapped in basis~\eqref{eq:qqbqqb-singlets-2}.

Perhaps more useful is the observation that ${\cal A}(Y)$ reduces to ${\cal
  A}^{\text{rigid}}(Y)$ identically if one approximates the $Y$-dependence of
${\cal G}$ by an ansatz that factorizes coordinate-dependence from
$Y$-dependence according to
\begin{align}
  \label{eq:GB-W-like-ansatz-gen}
  {\cal G}_{\bm x\bm y}(Y) \to f(Y)\, g(\bm x,\bm y)
\ ,
\end{align}
with arbitrary functions $f$ and $g$. The widely used Golec-Biernat-W\"usthoff
parametrization as well as the McLerran-Venugopalan model are of that
type. They replace ${\cal G}_{\bm x\bm y}(Y)$ by
\begin{align}
  \label{eq:GB-W-like-ansatz}
 & {\cal G}_{\bm x\bm y}^{\text{GB-W}}(Y)  =  
  Q_s^2(Y)\cdot (\bm x-\bm y)^2
\intertext{or}
 & {\cal G}_{\bm x\bm y}^{\text{MV}}(Y)  =  
   Q_s^2(Y)\cdot (\bm x-\bm y)^2 \ln[(\bm x-\bm y)^2\ \Lambda^2]
\end{align}
respectively. The factorization property~\eqref{eq:GB-W-like-ansatz-gen}
implies that the $Y$-dependence factors out of $a_Y$, $b_Y$, and $c_Y$;
eigenvectors of~\eqref{eq:qqb2-channel-matrix} become $Y$-independent and
$Y$-ordering is no longer an issue. The $(1,1)$-component of ${\cal
  A}^{\text{rigid}}(Y)$ in \eqref{eq:Exp_A_Y} is indeed identical to the
result obtained in \cite{Dominguez:2008aa} for the $\langle\Hat S_{\bm y'\bm
  x'}^{q\Bar q}\Hat S_{\bm x\bm y}^{q\Bar q}\rangle_Y$ correlator, where the
McLerran-Venugopalan model was employed. The 4-point correlator calculated in
\cite{Blaizot:2004wv} can also be recovered using a different color basis.

While this type of factorized ansatz is not compatible with the GT-evolution
equation~\eqref{eq:tilde-G-evo-short}, it has met with much phenomenological
success as an ansatz for the initial condition to evolution for dipoles,
sometimes supplemented with a short evolution interval to modify the initial
condition away from the simple models before one starts comparison with data.

Exploiting evolution to at least partially erase features of the initial
condition is useful in situations in which one expects most gluons in the
system to be perturbatively produced. In the process details of the initial
condition are erased and supplemented by universal properties imposed by the
nonlinearities of the evolution equation as the solutions of that equation
approach the asymptotic scaling regime.  For this situation, an analogous
strategy can also applied: Aiming at a phenomenological
comparison with data at $Y > Y_0$ one uses
\begin{align}
  \label{eq:ansatz-qqb2-initial}
  {\cal A}(Y) := P_Y \exp\biggl[-\int\limits_{Y_0-\Delta Y}^Y dY' 
\ {\cal M}(Y')
\biggr] 
\cdot
{\cal A}^{\text{rigid}}(Y_0-\Delta Y)
\end{align}
to parametrize the evolution of the four point function.

For sufficiently large $\Delta Y$, this procedure should erase much of the
arbitrariness introduced into ${\cal A}(Y_0)$ (and ${\cal A}(Y)$ at $Y > Y_0$)
by using any reference to ${\cal A}^{\text{rigid}}$ at all. In this sense,
Eq.~\eqref{eq:ansatz-qqb2-initial} should provide a reasonable \emph{ansatz}
for the $1/N_c$-corrections characteristic of the Gaussian truncation also
away from the coincidence limits (where they are correctly implemented by
construction). The quality of the result should improve with $Y$.

The strategy to calculate a correlator in the Gaussian truncation, as laid out
in the above, is in fact completely general. For any combination of quarks,
anti-quarks and gluons, one first needs to determine a full set of
non-equivalent singlet projections (if non-equivalent, their basis elements
will automatically be orthogonal to each other). From this one forms the
analogue of ${\cal A}(Y)$ in Eq.~\eqref{eq:qqb2-channel-matrix} as a
correlator matrix. The next step is to find ${\cal M}(Y)$ by expanding the
correlators to lowest order in ${\cal G}'$. With these ingredients one can
then construct (numerically) ${\cal A}(Y)$ via~\eqref{eq:ansatz-qqb2-initial},
using the solutions to~\eqref{eq:tilde-G-evo-short} as input.

The dimensionality of the problem generically grows with the number of quarks
and gluons: As already mentioned above, the $(q\Bar q)^2 g$-correlator gives
rise to 6-nonequivalent singlets and their transition elements. Adding more
Wilson lines will generically increase the number of singlets even more.  The
main exception to this rule are the baryon- or anti-baryon-correlators which
allow only one singlet to be formed via the totally antisymmetric tensor in
$N_c$-dimensions~\cite{Kovchegov:2001ni, Fukushima:2007dy}. As a consequence
the baryon correlators can be evaluated by direct exponentiation to give
\begin{align}
  \label{eq:GT-proton}
\langle \frac{\epsilon_{j_1\ldots j_{N_c}}}{N_c!} 
  [U_{\bm x_1}]_{j_1 i_1}
  \cdots 
  [U_{\bm x_{N_c}}]_{j_{N_c} i_{N_c}}
  \frac{\epsilon_{i_1\ldots i_{N_c}}}{N_c!}  
  \rangle(Y)
& =
\langle \frac{\epsilon_{j_1\ldots j_{N_c}}}{N_c!}  
  [U_{\bm x_1}^\dagger]_{j_1 i_1}
  \cdots 
  [U_{\bm x_{N_c}}^\dagger]_{j_{N_c} i_{N_c}}
  \frac{ \epsilon_{i_1\ldots i_{N_c}} }{N_c!} 
  \rangle(Y)
\notag \\
& =
  e^{
  -\frac{N_c+1}{2 N_c}\sum\limits_{i\neq j}^{N_c} {\cal G}_{Y;x_i,x_j} }
\ .
\end{align}

\section{Measuring correlator-factorization violations with target diffractive dissociation}
\label{sec:meas-corr-fact-directly}

As already indicated in Sec.~\ref{sec:Wilson-lines-beyond-total-cross}, taking
the difference of inclusive and exclusive vector-meson production
cross-sections, Eqns.~\eqref{eq:vm-inclusive-zeroeth}
and~\eqref{eq:vm-exclusive-zeroeth}, one directly probes correlator
factorization violations~\eqref{eq:corr-fact-viol}, since all but the
\tol{out}-out-overlap contributions cancel and one is left with
\begin{align}
  \label{eq:vm-inclusive-exclusive-diff-zeroeth}
  \diagram[height=1.1cm]{qqbint-L}
  \Biggl[
  \diagram[height=1.1cm]{vmwf-R}
  \cut
  \diagram[height=1.1cm]{vmwf-L}
  -
  \diagram[height=1.1cm]{vmwf-t-R}
  \cut
  \diagram[height=1.1cm]{vmwf-t-L}
  \Biggr]
  \diagram[height=1.1cm]{qqbint-R}
\ .
\end{align}

Using $t=-\bm l^2$, one finds the following expressions for
target-dissociating parts of the transverse and longitudinal cross-sections:
\begin{align}
  \label{eq:diff-diss}
  4\pi \frac{d\sigma_{T,L}}{d t} = &
  \int\limits_0^1 d\alpha d\alpha' \int d^2x d^2x' d^2y d^2 y'
  e^{-i \bm l\cdot[
    (\alpha\bm x+(1-\alpha)\bm y)
    -(\alpha'\bm x'+(1-\alpha')\bm y')
    ]
  }
\notag \\ & {}\times
  \Psi_{T,L}^*(\alpha',\bm x'-\bm y',Q^2)
  \Psi_{T,L}(\alpha,\bm x-\bm y,Q^2)
  \notag \\ & {}\times
  \biggl[\langle  \tr(U_{\bm y'} U_{\bm x'}^\dagger) 
                          \tr(U_{\bm x} U_{\bm y}^\dagger)
         \rangle_Y
         - \langle \tr(U_{\bm y'} U_{\bm x'}^\dagger) 
         \rangle_Y \langle
                 \tr(U_{\bm x} U_{\bm y}^\dagger)
         \rangle_Y
  \biggr]/N_c^2\ .
\end{align}
Let us first discuss a subtlety of this formula. The fully differential
vector-meson production cross-sections also depend on the fraction $z$ of the
meson longitudinal momentum with respect to the photon. This dependence has
been integrated out in~\eqref{eq:diff-diss}, in the regime where the eikonal
approximation we are using to describe the $\gamma^* A$ scattering is valid,
meaning with $1-\epsilon<z<1$ (the result does not depend on $\epsilon$ but
$\epsilon$ should be small for the result to hold). Over this kinematical $z$
range, even inclusive events feature a rapidity gap in the final state,
between the vector-meson and the system $X,$ of invariant mass $M_X,$ coming
from the dissociation of the target $A$: indeed from kinematics a small
$\epsilon$ implies $M_X\gtrsim M_A.$ The difference with exclusive events is,
that in these the target $A$ really escapes the collision intact. Obtaining
the $z$ dependence of the fully differential vector-meson cross-sections for
arbitrary $z,$ as well as a non-trivial $\epsilon$ (or maximum-$M_X$)
dependence of the $z$ integrated cross-section~\eqref{eq:diff-diss}, implies
to go beyond the eikonal approximation.

Using the change of variables $\bm r=\bm x-\bm y$, $\bm b=\alpha\bm
x+(1-\alpha)\bm y$ and $\bm r'=\bm x'-\bm y'$, $\bm b'=\alpha'\bm
x'+(1-\alpha')\bm y'$ allows to write the $t-$integrated cross-sections as
\begin{align}
  \label{eq:tot-diff-diss}
  \sigma_{T,L} = &
 \int\limits_0^1 d\alpha d\alpha' \int d^2r d^2 r'
  \Psi_{T,L}^*(\alpha',\bm r',Q^2)
  \Psi_{T,L}(\alpha,\bm r,Q^2)
  \notag \\ & {}\times
  \int d^2b
  \biggl[\langle  \tr(U_{\bm y'} U_{\bm x'}^\dagger) 
                          \tr(U_{\bm x} U_{\bm y}^\dagger)
         \rangle_Y
         - \langle \tr(U_{\bm y'} U_{\bm x'}^\dagger) 
         \rangle_Y \langle
                 \tr(U_{\bm x} U_{\bm y}^\dagger)
         \rangle_Y
  \biggr]/N_c^2
\ ,
\end{align}
where in the target averages $\langle\ldots\rangle_Y$, $\bm x=\bm
b+(1-\alpha)\bm r$, $\bm y=\bm b-\alpha\bm r$, and $\bm x'=\bm
b'+(1-\alpha')\bm r'$, $\bm y'=\bm b'-\alpha'\bm r'$.

These quantities directly probe corrections beyond the leading $1/N_c$
approximation of the BK truncation of JIMWLK evolution. The Gaussian
truncation, however, offers consistent finite expressions for such correlator
differences, that can be constructed from the solution to the evolution
equation for the Gaussian truncation~\eqref{eq:tilde-G-evo-short}. The
unfactorized term follows numerically from the $(1,1)$-component
of~\eqref{eq:ansatz-qqb2-initial}, the factors in the factorized term can be
read off from Eq.~\eqref{eq:UUdaggersol-app} to yield $\exp\bigl[-C_f\bigl(
{\cal G}_{\bm y'\bm x'}+ {\cal G}_{\bm x\bm y}\bigr)\bigr]$.

After $t$ integration, also heavy meson production cross-sections are of
particular interest: In the phase space region where the four Wilson line
correlator is dominated by the conicidence limit of the coordinates of the
heavy quark in amplitude and complex conjugate amplitude ($\bm x'=\bm x$), the
Gaussian truncation provides an analytical expression in terms of ${\cal G}$:
\begin{align}
  \label{eq:heavy-quark-conicidence-difference}
  \frac{\langle  \tr(U_{\bm y'} U_{\bm x}^\dagger) 
                          \tr(U_{\bm x} U_{\bm y}^\dagger)
         \rangle_Y}{N_c^2}
         = \frac{2 C_f}{N_c}e^{-\frac{N_c}2\bigl(
           {\cal G}_{Y,\bm y'\bm x}+
           {\cal G}_{Y,\bm x\bm y}-
           {\cal G}_{Y,\bm y'\bm y}\bigr)-C_f  {\cal G}_{Y,\bm y'\bm y}}
         +\frac{1}{N_c^2}e^{-C_f{\cal G}_{Y,\bm y'\bm y}}
\ .
\end{align}

This set of observables provides a very attractive tool to test nontrivial
features of JIMWLK-evolution in direct comparison with data. In the case of
vector-meson production, the cross-section difference we are interested in
could be measured at a future electron-ion collider~\cite{Deshpande:2005wd}.
In the case of heavy meson production, one may not be able to measure the
cross-sections directly.  Perhaps the inclusive cross-section can be extracted
in heavy-ion collisions, from studying the propagation of mesons through
cold-nuclear matter. In this case the Gaussian truncation provides $1/N_c^2$
corrections to the BK result.

To compare with data, however, one must find a way to treat the $b$-dependence
in Eqns.~\eqref{eq:diff-diss} and~\eqref{eq:tot-diff-diss} or any other meson
production cross-section.  JIMWLK evolution already fails to correctly
describe the $b$-dependence of the total cross-section
formula~\eqref{eq:dipole-cross}, since gluon emission in JIMWLK is
perturbative and has $\frac1{|\bm z|^2}$ power law tails that lead to
exponential growth of an initially finite size target. Comparison with data
already in this case is done by modeling the $b$-dependence using a factorized
$b$-profile according to
\begin{align}
  \label{eq:dis-profile}
  \sigma^{\text{dipole}}(\bm r^2,Y) = \int d^2b N_{\bm x\bm y;Y} 
  \to N_{\bm r^2;Y} \int d^2b\ T(\bm b^2)
\ .
\end{align}
$N_{\bm r^2;Y}$ is then taken from JIMWLK (or one of its truncations) while
the profile choice leaves only a normalization constant
$\int d^2b\ T(\bm b^2) \approx 2\pi R^2_{\text{target}}$.

The situation for the factorization violation measurements as well as the
separate inclusive and exclusive meson production cross-sections is more
sensitive to profiles: They explicitly enter the integrands of expressions
such as~\eqref{eq:vm-exclusive-zeroth-analytical} and will affect and distort
how $Y$-dependence as calculated from evolution equations is mapped onto the
actual behavior of cross-sections. Still, there are models on the market that
have been successfully applied to data (see for
instance~\cite{Kowalski:2006hc}) which one may use as starting point for
measuring correlator factorization violations, as in~\cite{Marquet:2009vs}. A
thorough evaluation of sensitivity on model details and model independent
statements will be left to a future phenomenological study.

\section{Conclusions}
\label{sec:conclusions}

We have shown that the high energy limit of inclusive vector-meson production
cross-sections in DIS is determined by two- and four-point Wilson-line
correlators [Eqns.~\eqref{eq:vm-inclusive-zeroeth},
and~\eqref{eq:vm-inclusive-zeroth-analytical}], whose energy dependence in
that limit is entirely determined by JIMWLK
evolution~\eqref{eq:vm-inclusive-evo}).  The situation for exclusive meson
production is different only in the contribution with four Wilson-lines, which
are factorized into a product of two point functions
(Eqns.~(\ref{eq:vm-exclusive-zeroeth}), and~(\ref{eq:qqb2-corr-gen-fact}))
with accordingly factorized (independent) JIMWLK
evolution~(\ref{eq:vm-exclusive-evo}).

Unlike particle multiplicities or forward particle production cross-sections,
only a minimal set of phenomenological assumptions is needed to compare with
data -- modelling is needed only with respect to impact parameter profiles and
meson wave functions. No additional assumptions about the applicability of
$k_t$-factorization or the twist expansion are needed as long as the energies
are high enough so that the no-recoil approximation remains valid to justify a
description solely in terms of Wilson line degrees of freedom. 

Already the inclusive and exclusive vector-meson production cross-sections
probe new information about JIMWLK evolution not visible in the total
cross-section. This is induced by their sensitivity to the energy dependence
of four point correlators (which cancels in simpler observables) both at
finite momentum transfer \emph{and} for the $t$-integrated cross-section (the
latter being both easier to calculate and to measure).

The difference of the two cross-sections --the target dissociative
contributions to vector-meson production-- is directly driven by the
correlator difference $\langle\Hat S_{\bm y'\bm x'}^{q\Bar q}\Hat S_{\bm x\bm
  y}^{q\Bar q}\rangle_Y -\langle \Hat S_{\bm y'\bm x'}^{q\Bar
  q}\rangle_Y\langle\Hat S_{\bm x\bm y}^{q\Bar q}\rangle_Y$. This quantity,
although expected to be small, is sensitive to entirely unexplored
contributions which are not accessible in the large-$N_c$ limit and even at
low densities start only at higher twist, with four gluons in the $t$-channel
necessary to yield a non-vanishing contribution.

We have explained how to efficiently implement JIMWLK-evolution of all the
correlators involved \emph{beyond} the large-$N_c$ limit using the Gaussian
truncation~\eqref{eq:func-Gaussian-average}. Such a step beyond the BK
approximation to JIMWLK-evolution is mandatory, since the BK approximation is
built on the large-$N_c$ limit and assumes $\langle\Hat S_{\bm y'\bm
  x'}^{q\Bar q}\Hat S_{\bm x\bm y}^{q\Bar q}\rangle_Y= \langle \Hat S_{\bm
  y'\bm x'}^{q\Bar q}\rangle_Y\langle\Hat S_{\bm x\bm y}^{q\Bar q}\rangle_Y$
from the outset. As a consequence, differences between inclusive- and
exclusive-vector-meson production become inaccessible -- they are beyond the
scope of the approximation. The GT approximation, on the other hand, allows to
calculate this difference and to obtain a finite result for the
target-dissociating part of the cross-section \eqref{eq:diff-diss}.

The GT approximation can be applied to even more general correlators -- the
method to obtain the four point correlator from a matrix evolution equation
outlined in Sec.~\ref{sec:quadr-evol-gauss} readily generalizes to arbitrary
$n$-point functions -- at the price of increasing numerical cost with each
added point. Constraints on the coordinate hyper-volumes help mitigate that
cost -- $t$-integrated cross-sections, for example, require less numerical
effort already at the four-point level.

While we have been mainly concerned with vector-meson production in DIS, there
exist other observables in hadron-hadron collisions known to involve
complicated correlators, to which our method can be applied. Heavy
quark-antiquark pair production~\cite{Fujii:2006ab} and inclusive dijet
production~\cite{Marquet:2007vb, Marquet:2008vm} which have already been
studied in the BK approximation can be reexamined in the Gaussian truncation.

The rich structure of the full JIMWLK equation and the Balitsky hierarchies
beyond the simplest truncations contain an immense amount of information that
can only be studied by widening the pool of observables considered.

\subsection*{Acknowledgments} 

We would like to thank Robi Peschanski for reading the manuscript and making
very useful comments.  C.M. wishes to thank the Department of Physical
Sciences of the University of Oulu, for hospitality while part of this work
was being completed.  C.M. is supported by the European Commission under the
FP6 program, contract No. MOIF-CT-2006-039860.  H.W. is supported by a
research grant of the University of Oulu.

\appendix

\section{Photon-vector-meson wave function overlaps}
\label{sec:photon-vector-meson-wave-function-overlaps}

Here we list the overlaps $\Psi_{T,L}(\alpha,\bm r,Q^2)$ of virtual photon and
vector-mesons introduced in Sec.~\ref{sec:evolution-dis-cross} and used
throughout.

The general structure and the complete photon-side contributions are
calculated from field theory, the vector-meson side relies on phenomenological
models for the functions $\phi_{L,T}$ entering below.

Defining $Q_f^2:=\alpha(1-\alpha)Q^2+m_f^2$, one has
\begin{align}
  \label{eq:wave-function-overlap-def}
  \Psi_T(\alpha,\bm r,Q^2) = &
  \hat{e}_f \sqrt{\frac{\alpha_e}{4\pi}} \frac{N_c}{\pi}
  \left\{m_f^2 K_0(r Q_f)\phi_T(r,\alpha) 
  - [\alpha^2+(1-\alpha)^2] Q_f K_1(r Q_f) \partial_r\phi_T(r,\alpha) \right\}
\ ,\\
  \Psi_L(\alpha,\bm r,Q^2) = &
  \hat{e}_f \sqrt{\frac{\alpha_e}{4\pi}} \frac{N_c}{\pi} \, 
  2Q\alpha(1-\alpha)K_0(r Q_f)  
  \left[M_V\alpha(1-\alpha)\phi_L(r,\alpha)
   +\delta\frac{m_f^2-\nabla_r^2}{M_V}\phi_L(r,\alpha)\right]
\ ,
\end{align}
separately for transversely and longitudinally polarized wave functions.

Among the models for $\phi_{L,T}$ used in the literature, two examples are of
particular interest: the {\em boosted Gaussian} (BG) wave functions
\cite{Nemchik:1994fp,Nemchik:1996cw}
\begin{equation}
  \phi^{BG}_{L,T} = N_{L,T}\,\exp\left[-\frac{m_f^2R^2}{8\alpha(1-\alpha)}+\frac{m_f^2R^2}{2}-\frac{2\alpha(1-\alpha)r^2}{R^2}\right]
\ ,
\end{equation}
used with $\delta=1$, and the {\em light-cone Gauss} (LCG) wave functions
\cite{Dosch:1996ss,Kulzinger:1998hw}
\begin{align}
  \phi^{LCG}_L = & N_L \,\exp\left[-r^2/(2R_L^2)\right]
\ ,\\
  \phi^{LCG}_T = & N_T\,\alpha(1-\alpha)\,\exp\left[-r^2/(2R_T^2)\right]
\ ,
\end{align}
used with $\delta=0$.  The parameters $R$ and $N_{L,T}$ are constrained by the
normalizations of the wave functions, as well as by electronic decay
widths. They are given for instance in the Appendix of
Ref.~\cite{Marquet:2007qa}, along with the effective charges $\hat{e}_f$,
quark masses $m_f$, and meson masses $M_V$.

\section{Expressing the Gaussian truncation in terms of a non-local 
Gaussian distribution of $\rho$'s}
\label{sec:link-with-non-local-gauss}

In this Appendix, we show that the GT approximation is equivalent to using the
following ansatz for the CGC wave function in terms of target color sources
$\rho$:
\begin{equation}
  \label{eq:ansatz}
W_Y[\rho]=\exp\left(-\int\limits^Y dY' 
  \int\! d^2x\, d^2y\  \frac{\rho_c(\bm x,Y')\rho_c(\bm y,Y')}
{2\mu^2(Y',\bm{x},\bm{y})}\right)\ .
\end{equation}
It is a Gaussian distribution for the color sources $\rho_c,$ whose variance
$\mu^2$ represents the transverse color charge density squared on the path of
a projectile moving along the $x^-$ direction. The longitudinal extend of the
target, over which $\mu^2$ doesn't vanish, is proportional to $e^{-Y}.$ Rather
than labeling $\mu^2$ with this explicit $Y$-dependence, it is recast in the
boundaries of the rapidity integration (equivalent to an $x^-$ integration).

In Eq.~\eqref{eq:target-average}, after expanding the Wilson-line content of
``$\ldots$'' in powers of the color field $A^+$, (we work in the light-cone
gauge $A^-=0$ where $A^+$ is a linear function of $\rho$), any average with
the Gaussian weight $W_Y[\rho]$ can be computed using Wick's theorem with
$\mu^2(Y',\bm x,\bm y)$ parametrizing the $\rho$-correlators
\begin{equation}
\langle\rho_c(\bm x,Y')\rho_d(\bm y,\bar{Y}')\rangle_Y=
\delta_{cd}\delta(Y'-\bar{Y}')\mu^2(Y',\bm x,\bm y)
\ .
\end{equation} 

To establish the connection with~\eqref{eq:func-Gaussian-average}, one first
notes that it simply restates Wick's theorem as
\begin{equation}
  \langle \ldots \rangle_Y = \exp\biggl\{ 
  \int\limits^Y dY'd\bar{Y}' \int\! d^2x\, d^2y\ 
   \langle A^{c +}_{\bm x,Y'} A^{d +}_{\bm y,\bar{Y}'}\rangle_Y
\frac{\delta}{\delta A^{c +}_{\bm x,Y'}}
    \frac{\delta}{\delta A^{d +}_{\bm y,\bar{Y}'}}
  \biggr\} \ldots
\ ,
\end{equation}
and assumes that the $AA$ correlator is in fact local in $Y$
\begin{equation}
\langle A^{c +}_{\bm x,Y'} A^{d +}_{\bm y,\bar{Y}'}\rangle_Y
=
\delta_{cd}\delta(Y'-\bar{Y}')G_{Y',\bm{x y}}\ ,
\end{equation}
to pick the leading contribution in the non-abelian exponentiation
theorem.

The only step left is the translation of the $AA$-correlator into
$\rho$-correlators via the Yang-Mills equation $-\nabla_\perp^2 A^{c
  +}=\rho_c$. Inverting this relation gives
\begin{equation}
A^{c +}_{\bm x,Y'} 
=
\int d^2y\ 
G_0(\bm x-\bm y)\rho_c(\bm y,Y')
\ ,\quad 
G_0(\bm x)
=
\int\frac{d^2k}{{(2\pi)^2}} \frac{e^{i\bm k\cdot\bm x}}{\bm k^2}\ ,
\end{equation}
and therefore
\begin{equation}
G_{Y',\bm{x y}} = \int d^2zd^2z'\ 
\mu^2(Y',\bm z,\bm z')
G_0(\bm{x}-\bm{z})G_0(\bm{z}'-\bm{y})
\ .
\end{equation}

\section{Decomposing the four-Wilson-line correlator as $q^2\Bar q^2$}
\label{sec:decomp-four-wils-as-q2qb2}

This employs symmetrizers and
antisymmetrizers defined as
\begin{align*}
  \diagram[height=.65cm]{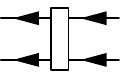}
  := \frac12\Bigl[
  \diagram[height=.65cm]{q2id}
  +
  \diagram[height=.65cm]{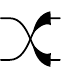}
  \Bigr] 
\hspace{1cm}\text{and} \hspace{1cm} 
   \diagram[height=.65cm]{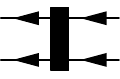}
   := \frac12\Bigl[
  \diagram[height=.65cm]{q2id}
  -
  \diagram[height=.65cm]{q2swap}
  \Bigr] 
\end{align*}
and the decomposition of $q q$  and $\Bar q\Bar q$ states 
according to
\begin{align*}
  \diagram[height=.65cm]{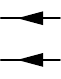}  
  = 
  \diagram[height=.65cm]{qqsymm}
  +
  \diagram[height=.65cm]{qqasymm} 
\hspace{1cm}\text{and} \hspace{1cm} 
\diagram[height=.65cm]{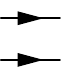}  
  = 
  \diagram[height=.65cm]{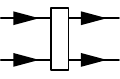}
  +
  \diagram[height=.65cm]{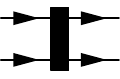}
\end{align*}
to define a basis for two non-equivalent singlet channels as
\begin{equation}
  \label{eq:q2qb2-basis}
  \frac1{\diagram[height=.7cm]{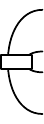}\diagram[height=.7cm]{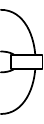}^{\frac12}}
  \diagram[height=.8cm]{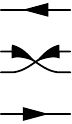}\diagram[height=.8cm]{symm-R} 
  \hspace{1cm}\text{and} \hspace{1cm}
  \frac1{\diagram[height=.7cm]{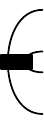}\diagram[height=.7cm]{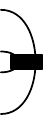}^{\frac12}}
  \diagram[height=.8cm]{qqbqqbtoqqqbqb}\diagram[height=.8cm]{asymm-R} 
 \ ,
\end{equation}
with normalization factors arising from
\begin{align*}
  \diagram[height=.8cm]{symm-L}\diagram[height=.8cm]{symm-R},
  \diagram[height=.8cm]{asymm-L}\diagram[height=.8cm]{asymm-R}
  \leftrightarrow
  \frac14\biggl[
  \diagram[height=.8cm]{qqqbqbthru-L}\pm\diagram[height=.8cm]{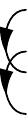}
  \biggr]
  \biggl[
  \diagram[height=.8cm]{qqqbqbthru-R}\pm\diagram[height=.8cm]{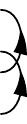}
  \biggr]
  =
  \frac14\biggl[
  \underset{N_c^2}{
    \diagram[height=.8cm]{qqqbqbthru-L}\diagram[height=.8cm]{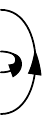}
    }
  \pm
  \underset{N_c^{\phantom{2}}
   }{
  \diagram[height=.8cm]{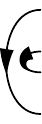}\diagram[height=.8cm]{qqqbqbmix-R}
  }
  \pm
  \underset{N_c^{\phantom{2}}
   }{
  \diagram[height=.8cm]{qqqbqbmix-L}\diagram[height=.8cm]{qqqbqbthru-R}
  }
  +
  \underset{N_c^2}{
  \diagram[height=.8cm]{qqqbqbmix-L}\diagram[height=.8cm]{qqqbqbmix-R}
  }
  \biggr]
  =\frac{N_c(N_c\pm 1)}2
\ .
\end{align*}
The off-diagonal elements of Eq.~\eqref{eq:qqb2-channel-matrix} in this last
basis~\eqref{eq:q2qb2-basis} vanish when $\bm y'\to \bm x$ or $\bm x'\to\bm y$,
since anti-symmetrization of a symmetric object yields zero
\begin{align}
  \label{eq:q2-or-qb2-transition-limit}
 \begin{smallmatrix}
    i_1 \\ i_2 
  \end{smallmatrix}
  &
  \diagram[height=.7cm]{qqsymm}
  \begin{smallmatrix}
    U_{\bm y'} \\ U_{\bm x} 
  \end{smallmatrix}
\diagram[height=.7cm]{qqasymm}
  \begin{smallmatrix}
    j_1 \\ j_2 
  \end{smallmatrix}
  \xrightarrow{\bm y'\to \bm x}  
\begin{smallmatrix}
    i_1 \\ i_2 
  \end{smallmatrix}
 \begin{smallmatrix}
    U_{\bm x} \\ U_{\bm x} 
  \end{smallmatrix} 
\diagram[height=.7cm]{qqsymm}
 \diagram[height=.7cm]{qqasymm}
  \begin{smallmatrix}
    j_1 \\ j_2 
  \end{smallmatrix}
= 0
\ .
\end{align}
The surviving correlators that appear on the diagonals in this basis in these
limits constitute the last set of coincidence limit conditions on ${\cal
  A}(Y_0)$. Using the basis~\eqref{eq:q2qb2-basis}, one obtains explicit
expressions for ${\cal A}$ in the limit $\bm x'=\bm y$
\begin{subequations}
\label{eq:symmasymm-diagonal-limits}
\begin{align}
  \label{eq:symmasymm-diagonal-limitsxp=y}
    \begin{pmatrix}
      e^{-(C_f+\frac{N_c-1}{2 N_c})(
        {\cal G}_{\bm x,\bm y}
        +
        {\cal G}_{\bm y,\bm y'}
        -
        {\cal G}_{\bm x,\bm y'}
        )
        -C_f{\cal G}_{\bm x,\bm y'}
      }
      & 0 \\ 0 &
      e^{-(C_f-\frac{N_c+1}{2 N_c})(
        {\cal G}_{\bm x,\bm y}
        +
        {\cal G}_{\bm y,\bm y'})
        -\frac{N_c+1}{2 N_c}{\cal G}_{\bm x,\bm y'}
      }
    \end{pmatrix}
\ ,
\intertext{while the limit $\bm y'=\bm x$ leads to}
  \label{eq:symmasymm-diagonal-limitsyp=x}    
   \begin{pmatrix}
      e^{-(C_f+\frac{N_c-1}{2 N_c})(
        {\cal G}_{\bm x,\bm y}
        +
        {\cal G}_{\bm x,\bm x'}
        -
        {\cal G}_{\bm x',\bm y}
        )
        -C_f{\cal G}_{\bm x',\bm y}
      }
      & 0 \\ 0 &
      e^{-(C_f-\frac{N_c+1}{2 N_c})(
        {\cal G}_{\bm x,\bm y}
        +
        {\cal G}_{\bm x,\bm x'})
        -\frac{N_c+1}{2 N_c}{\cal G}_{\bm x',\bm y}
      }
    \end{pmatrix}
\ .
\end{align}
\end{subequations}

These structures are not completely meaningless: For $N_c=3$ the
anti-symmetrized correlator (lower right corner) must agree with the color
correlator of a proton (anti-proton) Eq.~\eqref{eq:GT-proton}.  This is indeed
the case, since the prefactors of all three terms in the exponent become equal
for $N_c=3$:
\begin{equation*}
  \label{eq:n=3-proton}
  (C_f-\frac{N_c+1}{2 N_c})\xrightarrow{N_c\to 3} \frac23 
  \xleftarrow{N_c\to 3}\frac{N_c+1}{2 N_c}
 \ . 
\end{equation*}

The symmetrized channel shows a structure similar to the $q\Bar q g$
correlator with ($C_f+\frac{N_c-1}{2 N_c})$ replacing the $\frac{N_c}2$ present
there. The two agree at $N_c=2$.

\bibliography{master,morerefs}                   \bibliographystyle{JHEP}  

\end{document}